Erika De Francesco[1,3], Salvatore Iiritano [1],
Antonino Spagnolo[2], Marco Iannelli[2]

[1] Exeura s.r.l..
V. Pedro Alvares Cabrai
87036 Rende (CS)
{erika.defrancesco, salvatore.iiritano}@exeura.com

[2] I.F.M. s.r.l.
V. Lombardi
88100 Catanzaro
{ nino.spagnolo, marco.iannelli}@ifm.it

[3] Università della Calabria
Dipartimento di Matematica
V. P. Bucci
87036 Rende (CS)

# A methodology for semi-automatic classification schema building

# Una metodologia per la creazione semi-automatica di schemi di classificazione

**Abstract:**

This paper describe a methodology for semi-automatic classification schema definition (a classification schema is a taxonomy of categories useful for automatic document classification). The methodology is based on: (i) an extensional approach useful to create a *typology* starting from a document base, and (ii) an *intensional* approach to build the classification schema starting from the typology. The extensional approach uses clustering techniques to group together documents on the basis of a similarity measure, whereas the intensional approach uses different operations (aggregation, reduction, generalization specialization) to define classes.

**Abstract:**

In questo documento viene descritta una metodologia semi-automatica per la costruzione di schemi di classificazione, ovvero tassonomie di categorie utili nelle applicazioni di classificazione documentale. Partendo da un set documentale che rappresenta la base di conoscenza del dominio in esame, il metodo si basa su: (i) un approccio *estensionale* per la creazione di una *tipologia*, e (ii) approccio *intensionale* volto a definire la tassonomia delle categorie di riferimento.

L'approccio estensionale procede raggruppando i documenti di interesse in diversi sottoinsiemi secondo opportune misure di *similarità* calcolate sulla base di un insieme di proprietà o features. L'approccio intensionale, invece, sfrutta operazioni di aggregazione, riduzione, specializzazione e generalizzazione per la definizione delle classi.

# 1 Introduzione

L'aumento esponenziale delle informazioni gestite e pubblicate dai sistemi di document e content management determina un problema di *information overloading* (sovraccarico) che rende spesso impossibile arrivare al contenuto davvero rilevante per l'utente. All'accumulo indiscriminato di informazioni va contrapposta una gestione ragionata dei contenuti, che consenta di organizzare tutti i contenuti sulla base di categorie significative nel contesto di riferimento.

Le attività di classificazione hanno il fine di assegnare i documenti rispetto ad opportune categorie del dominio in esame in modo che possano essere presentati ai fruitori (e da questi possano essere reperiti nelle categorie a disposizione) servendosi di criteri riconducibili ad una certa razionalità, in certi casi arrivando alla possibilità di avvalersi di regole precise e di procedure.

Dal punto di visto operativo, l'esecuzione di un'attività di classificazione prevede due macro-attività:

1) definizione di una *ontologia*, intesa come l'insieme delle categorie di interesse e delle relazioni tra le stesse, che possono essere sia di tipo gerarchico (relazioni di *specializzazione*), che di tipo non gerarchico;

2) definizione di un insieme di *criteri* di assegnamento dei documenti alle classi individuate. Con riferimento alla realizzazione di strumenti automatici, tali criteri si possono sostanziare in istruzioni *hard-coded* all'interno di algoritmi o in *regole* applicate da sistemi general purpose.

## 2 Fundamentum divisionis

Il *fundamentum divisionis* rappresenta il meccanismo logico attraverso il quale un concetto generale viene suddiviso per definire concetti di classe specializzati rispetto ad una particolare proprietà.

Si consideri, ad esempio, il concetto di *sistema politico*, e si faccia riferimento alla proprietà rappresentata dalla *legittimazione dei governanti*: in questo caso è possibile definire le categorie seguenti:

- sistema politico teocratico;
- sistema politico autocratico;
- sistema politico aristocratico;
- sistema politico plutocratico;
- sistema politico democratico.

Mentre il *fundamentum divisionis* caratterizza la classificazione nel suo complesso perché definisce il *criterio di specializzazione* delle classi, il *livello di generalità* è una proprietà di ciascun singolo concetto di classe. Due concetti A e B sono allo stesso livello di generalità quando:

1. le categorie derivate da A non fanno parte dell'insieme delle categorie derivate da B, e viceversa;
2. le categorie derivate da A non fanno parte dell'insieme delle categorie derivate da un concetto C che sia allo stesso livello di generalità di B, e il simmetrico vale per B.

# 3 Tipologie e Ontologie per la classificazione

## 3.1 Tipologie

Una *tipologia* è un insieme di oggetti, definiti in questo caso *tipi*, ognuno dei quali è ottenuto considerando simultaneamente più *fundamenta divisionis*. Ogni *tipo* è caratterizzato dall'intero insieme di *proprietà* delle singole classi che lo costituiscono, indipendentemente dall'ordine in cui sono applicati i *fundamenta*. Ciò significa che il tipo delle persone di nazionalità italiana e medici di professione coincide con il tipo dei medici che hanno nazionalità italiana.

La costruzione di una tipologia è caratterizzata dal fatto che il numero dei tipi che la costituiscono (detto la 'potenza' della tipologia) aumenta esponenzialmente all'aumentare del numero di fundamenta divisionis applicati. Inoltre, generando i tipi tramite una combinazione dei fundamenta applicati, diventa molto probabile che molti tipi siano una mera possibilità logica, priva di interesse concettuale e che alcuni tipi non abbiano specializzazione o abbiano una specializzazione molto ridotta.

Questi limiti applicativi alla costruzione delle tipologie rendono necessaria un'operazione di 'riduzione dello spazio di attributi' che, minimizzando il numero di classi individuate per ogni *fundamentus*, comporta una riduzione della dimensione della tipologia, ed una conseguente maggiore espressività dei tipi ottenuti.

Quando, anche dopo una fase di riduzione dello spazio degli attributi, il numero di tipologie ottenute continua a rimanere eccessivamente alto, l'unico rimedio applicabile è invece la riduzione manuale del numero dei tipi, che non deve avvenire per eliminazione bensì per aggregazione di due o più tipi in uno solo, che abbia specializzazione più vasta. Il processo di aggregazione deve essere governato da considerazioni di prossimità semantica fra i tipi (alla luce degli scopi per cui è costruita la tipologia), moderate dall'opportunità di bilanciare la loro estensione. Non è opportuno fondere due tipi, quale che sia la loro prossimità semantica, se la specializzazione del tipo così ottenuto è tale da soverchiare le specializzazioni degli altri tipi.

Una tipologia è denominata *tassonomia* quando i diversi *fundamenta divisionis* sono considerati in successione. In questo caso l'ordine in cui i *fundamenta* vengono considerati ha importanza: la tassonomia che si produce usando la proprietà X per specializzare un concetto, e poi la proprietà Y per fare altrettanto con i concetti specializzati che si sono ottenuti, non è la stessa che si produce usando prima la proprietà Y e poi la X.

Nella specializzazione dei tipi è necessario tenere in considerazione le seguenti proprietà:

- i tipi specializzati devono essere tra loro mutuamente esclusivi;

- l’insieme dei tipi della tassonomia deve essere, nel complesso, esaustivo;
- si possono usare tipi residuali per garantire l'esaustività e comprimere il numero delle specie.

Le tassonomie prevedono il mantenimento di una struttura gerarchica: ciò implica che tutte le specializzazioni dello stesso tipo devono essere allo stesso livello di generalità. Ciò potrebbe comportare una proliferazione di tipi specializzati finalizzata alla necessità di raggiungere l'esaustività.

Tuttavia, questi inconvenienti possono essere aggirati con un abile impiego dei tipi residuali: se la specializzazione mediante un certo *fundamentum* produce troppi tipi specializzati, alcuni possono essere raggruppati in un tipo residuale, che potrà essere ulteriormente specializzato al livello successivo.

## 3.2 Ontologie per la classificazione

Le ontologie rappresentano il modo più naturale per definire le categorie di riferimento nell’ambito di un’attività di classificazione. Una ontologia è un insieme di *categorie* a diversi livelli di generalità, ottenuto a partire da una tipologia/tassonomia facendo riferimento ad un unico *fundamentum divisionis*. Il ricorso a classi a diverso livello di generalità è in genere giustificato per rimediare a gravi squilibri nel numero di documenti assegnabili delle classi allo stesso livello.

Immaginiamo di voler definire un’ontologia relativa alle confessioni religiose. In tal caso possiamo partire dalla tassonomia rappresentata nella figura seguente.

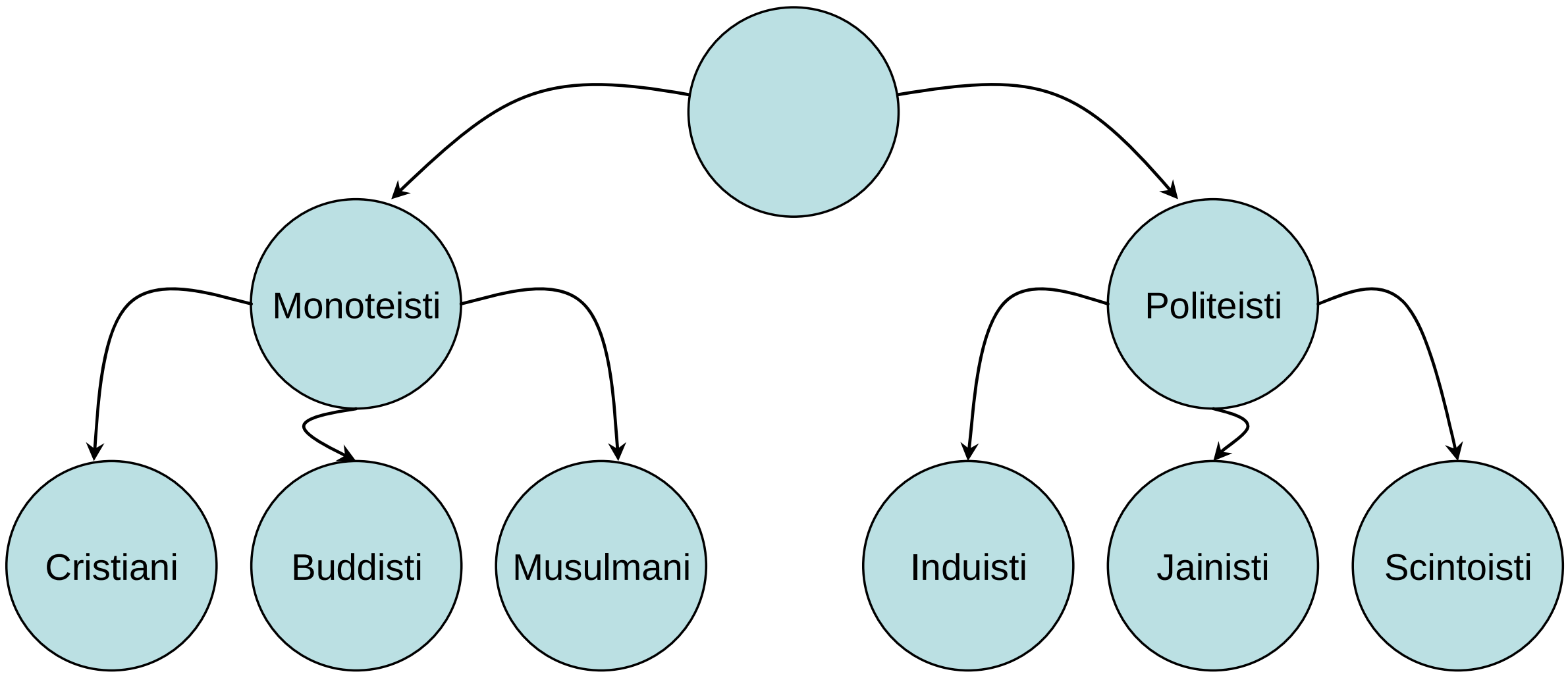

Obiettivo dell’ontologia è quello di definire l’insieme di tutte le possibili classi di professioni religiose, bilanciando in qualche modo il numero di elementi appartenenti alle classi, senza perdere l'esaustività.

A tal fine si può immaginare di eseguire le seguenti operazioni:

1) introdurre dei livelli più generali di classificazione, suddividendo tra *credenti* ed *atei*, e suddividendo ulteriormente i *credenti* in *credenti in divinità* ed *animisti*;

2) specializzare ulteriormente la classe *monoteisti* introducendo una classe residuale *altri monoteisti*;

3) specializzare ulteriormente la classe *politeisti* introducendo una classe residuale *altri politeisti*;

4) specializzare la classe *cristiani* introducendo la classe *cattolici*;

5) specializzare la classe *musulmani* introducendo la classe *sunniti*.

Lo schema finale si presenta come nella figura seguente.

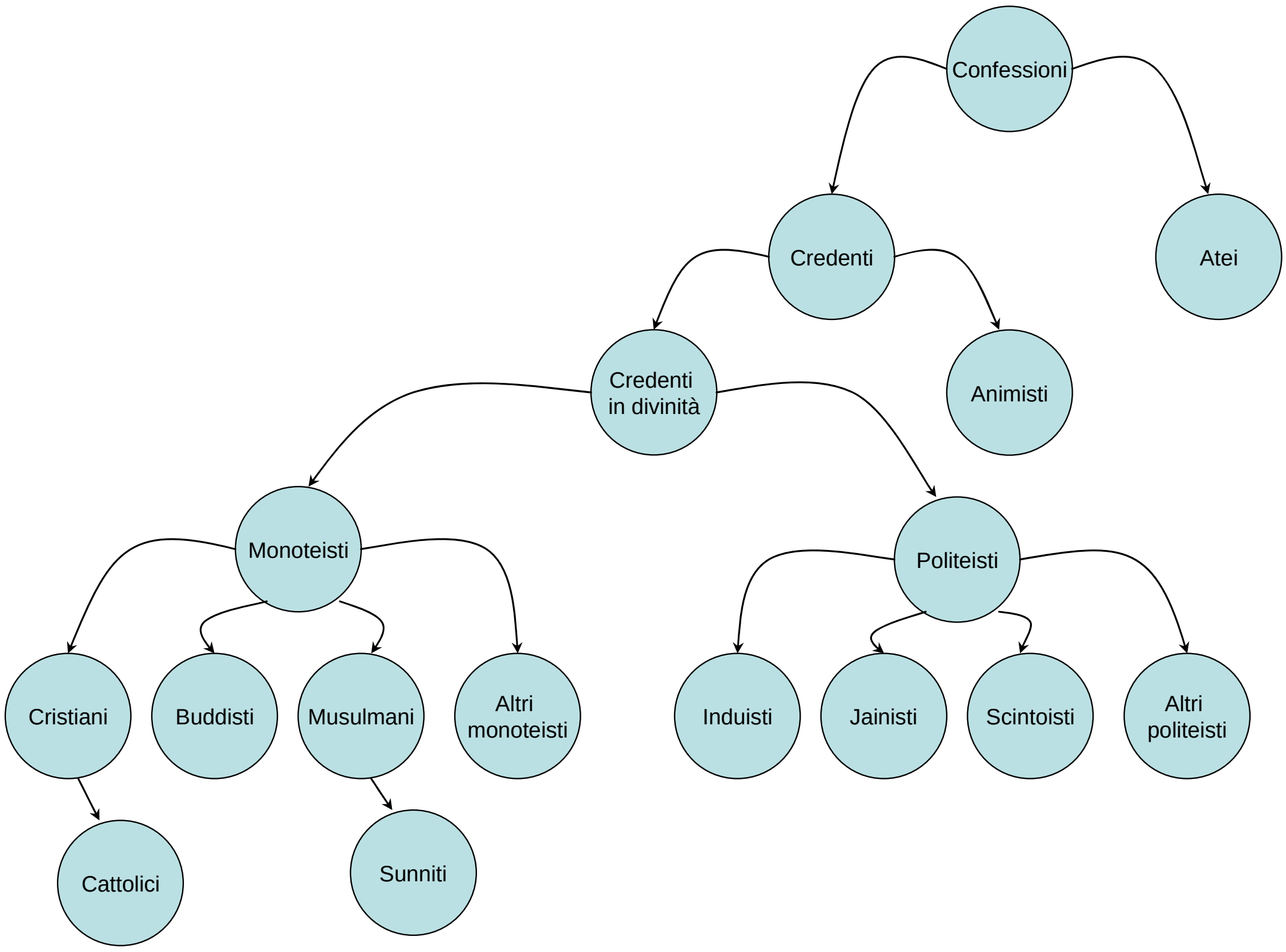

Tra le categorie di un’ontologia per la classificazione è possibile individuare due tipologie di legami, sulla base di particolari tipi di *fundamentum divisionis* e delle relative definizioni operative:

- *relazioni d'ordine*: se percepiamo gli stati su una proprietà come ordinati, è possibile riprodurre tale ordine nei rapporti fra le classi dello schema di classificazione. Supponiamo che uno stesso sistema educativo preveda tre livelli: elementare, medio e avanzato, e che il livello avanzato sia distinguibile in tre classi diverse (A, B, C). Si può costituire lo schema rappresentato nella figura seguente.

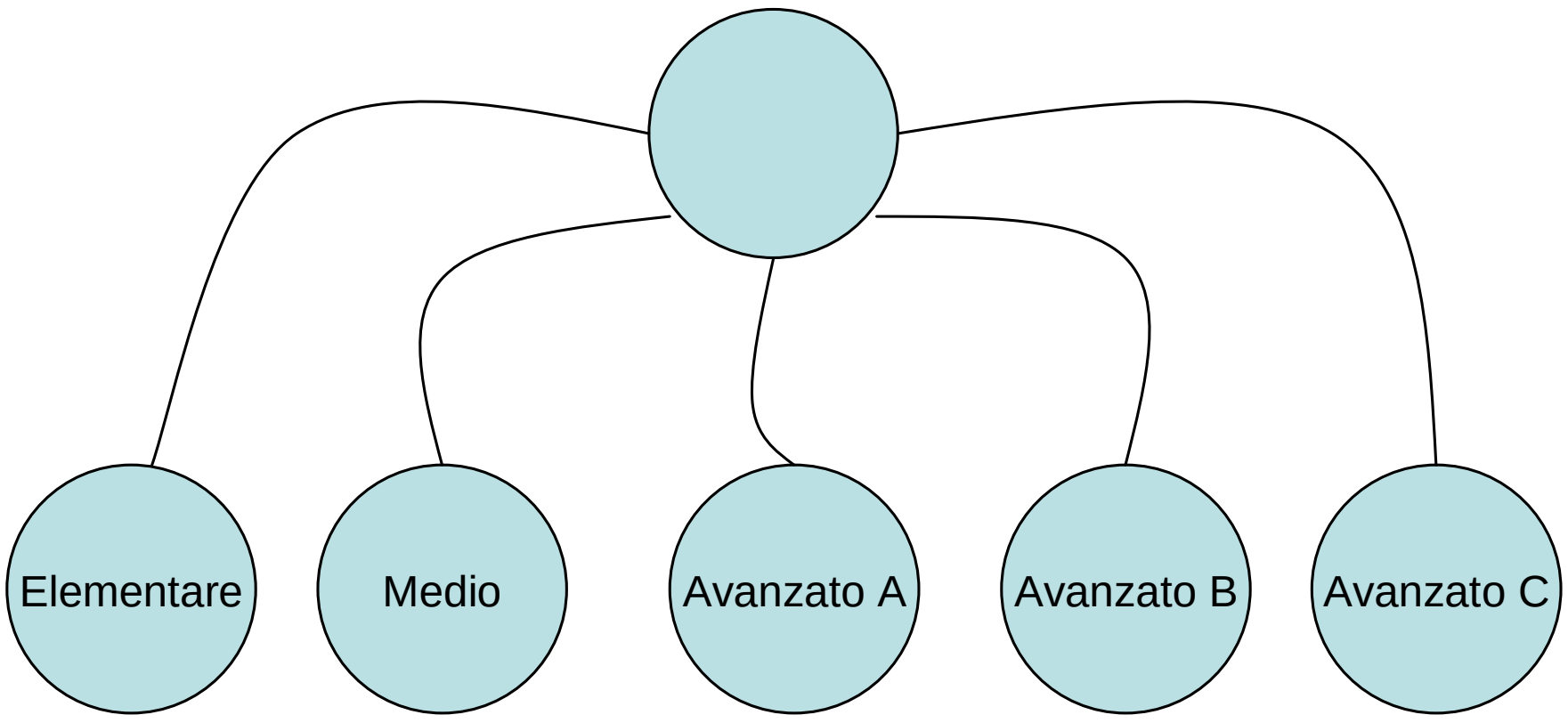

- *rapporti quantitativi*: gli stati di una proprietà possono anche esser percepiti come allineati lungo un *continuum*, cioè isomorfi ai numeri reali. Tuttavia, poiché non è possibile registrare numeri con infinite cifre, il *continuum* dovrà essere segmentato mediante un'unità di misura, o — se tale unità manca — mediante una procedura di *scaling*, come si fa di frequente nelle scienze sociali. In alternativa, possiamo percepire gli stati su una proprietà come accertabili mediante un conteggio. Contando, misurando, o applicando qualche forma di *scaling*, costituiamo automaticamente uno schema di classificazione, le cui classi possono essere: nessun peso / 1 grammo / 2 grammi..., oppure: nessun figlio/ 1 figlio / 2 figli..., oppure: approvo pienamente / approvo con riserva / sono incerto...

# 4 Metodologia per la costruzione di un'applicazione di classificazione

## 4.1 Definizione di ontologie

La definizione dell'ontologia per la classificazione si fonda sull'esecuzione di due attività:

1. *creazione di una tipologia*, attraverso un *approccio estensionale*: in questo caso si procede raggruppando i documenti di interesse in diversi sottoinsiemi a seconda di similarità calcolate sulla base di un insieme di proprietà (*features*). Una volta costruiti i gruppi di documenti, è possibile iterare l'attività di raggruppamento per ognuno dei gruppi creati;

2. *definizione dello schema*, attraverso un *approccio intensionale*: si procede manualmente alla definizione delle categorie di riferimento tramite operazioni di *aggregazione*, *riduzione*, *specializzazione* e *generalizzazione* delle classi individuate nella tipologia.

La figura seguente mostra il processo nel suo insieme:

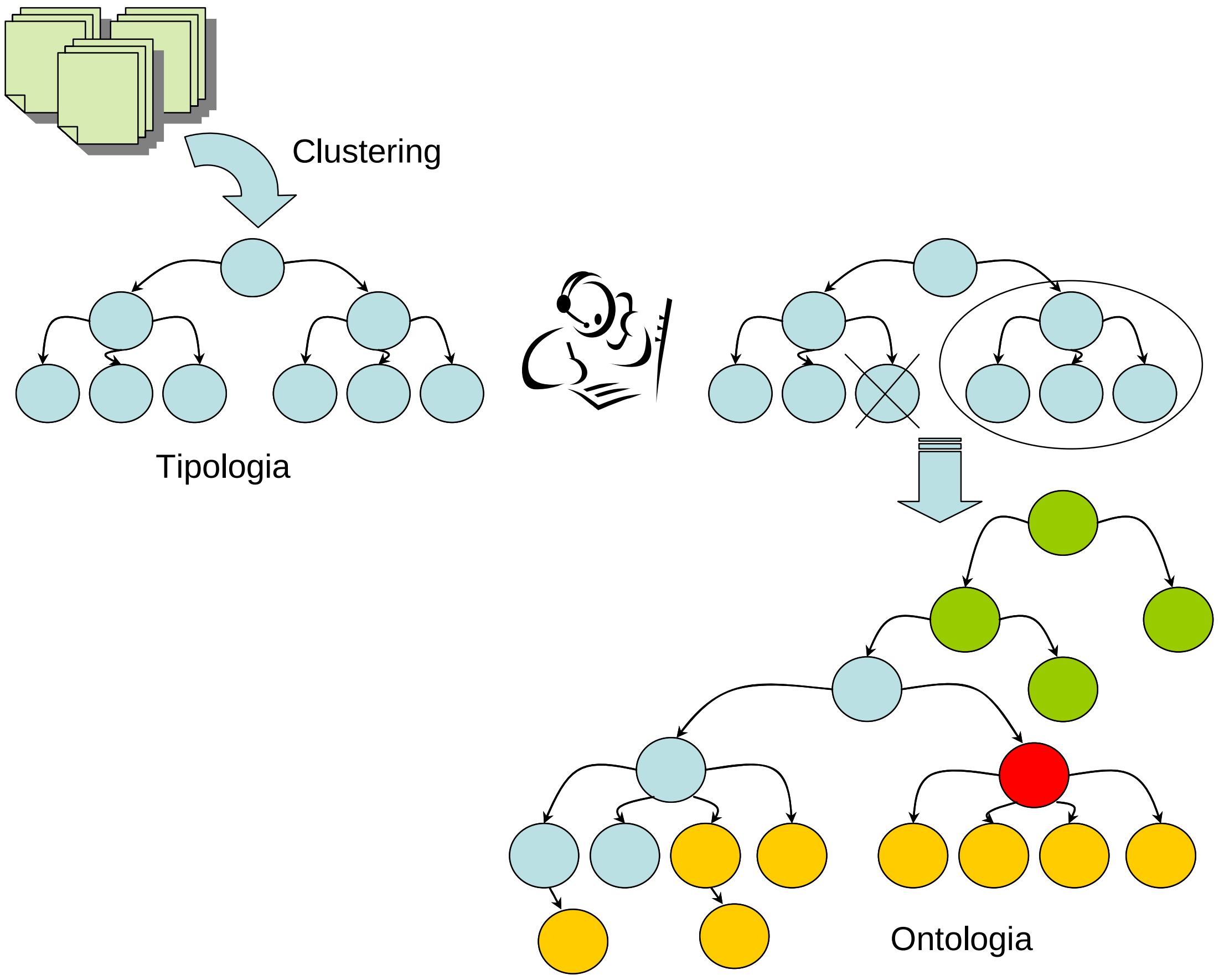

La tassonomia generata attraverso il clustering viene analizzata dall'utente che decide di eliminare la categoria contrassegnata dalla croce e di raggruppare le categorie contenute nel cerchio; l'ontologia finale contiene categorie aggiunte per generalizzazione (in verde), per specializzazione (arancio), oltre alla categoria ottenuta per aggregazione delle categorie da raggruppare (rosso).

### *4.1.1 Dai documenti alla tipologia: approccio estensionale*

La classificazione estensionale (o *cluster analysis*) è un insieme di operazioni attraverso le quali si raggruppano gli oggetti di un insieme in due o più sottoinsiemi in modo da massimizzare la somiglianza rispetto ad una serie di proprietà (*features*). Le features considerate sono abitualmente più di una, per cui il meccanismo classificatorio utilizza più *fondamenta divisionis* contemporaneamente: l'output dell'approccio estensionale, dunque, è una tipologia.

L'esecuzione di una classificazione estensionale per un insieme ampio di oggetti richiede:

- l'organizzazione delle features secondo una *matrice di attributi*, nella quale siano evidenziati:
  - gli oggetti sui quali operare la classificazione (cioè i vettori-riga nella matrice);
  - le proprietà i cui stati verranno presi in considerazione per massimizzare l'omogeneità interna e la mutua eterogeneità delle classi (cioè le variabili, o vettori-colonna della matrice);
- la definizione della "funzione distanza", ovvero di una formula logica e/o matematica per combinare le differenze sulle varie proprietà considerate in una singola "misura" della differenza fra due oggetti o eventi.

L'esecuzione di una classificazione estensionale pone il problema di definire il numero di *features* usate per il raggruppamento: infatti, se da un lato si tende a ridurre questo numero per agevolare le prestazioni dei sistemi automatici di clustering, dall'altro aumentando il numero di variabili cresce la probabilità di una corretta classificazione, perché tralasciando alcune proprietà si incorre in una perdita di informazione di entità ignota, poiché lavorando sulla matrice dei dati non c'è modo di considerare informazioni non incluse nella matrice stessa.

Ciò implica la necessità di stabilire opportuni criteri di esclusione di *features* non abbastanza discriminanti, o che non soddisfano altri requisiti, oltre a criteri per ponderare proprietà considerate più importanti. Inoltre, nella definizione delle classi, è necessario ulteriormente stabilire se il valore calcolato della funzione distanza debba essere "corretto" da criteri che tengano in considerazione (i) differenze intraclasse massime e/o interclasse minime su proprietà specifiche e (ii) limiti massimi e/o minimi alla numerosità, assoluta e/o proporzionale, di una classe.

Una volta eseguito un primo raggruppamento degli oggetti nelle varie classi, la classificazione estensionale può prevedere un ulteriore clustering all'interno dei singoli gruppi: in questo caso il risultato sarà rappresentato da una tassonomia, con tanti livelli quanti sono i raggruppamenti successivi che vengono eseguiti.

### *4.1.2 Dalla tipologia all'ontologia: approccio intensionale*

Nella sua forma strutturalmente più semplice, la classificazione intensionale può essere considerata un processo di esplicazione o chiarificazione concettuale. Il concetto relativo ad una classe viene formato o chiarificato mediante la definizione dei suoi confini semantici con i concetti relativi alle altre classi.

La classificazione estensionale sfrutta un sistema di attributi (metadati) mutuamente esclusivi rappresentanti ciascuno un aspetto o proprietà persistente dell'oggetto e capaci – nel loro insieme – di descrivere esaustivamente l'oggetto stesso. Tali attributi sono contraddistinti da queste peculiarità:

- sono invariabili dal punto di vista semantico (ad es. la proprietà COLORE di un oggetto può variare in termini di valori che può assumere – giallo, rosso etc. – ma è invariabile come concetto; cioè quell'oggetto avrà sempre un colore)
- costituiscono un insieme aperto, per cui è sempre possibile aggiungere nuovi attributi nel processo di classificazione;
- sono utilizzabili come attributi di ricerca sia singolarmente sia in combinazione.

Tali caratteristiche rendono particolarmente efficace l'adozione di questo sistema in ambienti digitali, per un più veloce ed efficiente rinvenimento dell'informazione.

Nella dottrina classica le ontologie per la classificazione sono caratterizzate da due proprietà fondamentali:

- *mutua esclusività*: i confini fra le classi sono rigidamente delimitati, ovvero data una qualsiasi coppia di classi, nessun oggetto può essere attribuito ad entrambe le classi della coppia;
- *esaustività*: ogni possibile stato sulla proprietà che si è adottata come *fundamentum divisionis* deve essere stato assegnato ad una delle classi.

Considerate insieme, la mutua esclusività di ogni possibile coppia di classi e l'esaustività del loro complesso garantiscono che ogni referente del concetto di genere sia assegnato ad una ed una sola delle classi che lo specificano.

I requisiti della dottrina classica sono perfettamente ragionevoli, ed è abbastanza agevole rispettarli quando gli oggetti che si vogliono classificare sono pensati a tavolino. Nel caso reale, non sempre l'appartenenza di un oggetto ad una delle categorie arbitrariamente costituite è determinabile in modo univoco (sì o no); talvolta è una questione di grado, ed in generale la mutua esclusività di ogni coppia di classi è un obiettivo che viene conseguito solo stabilendo delle regole irrealisticamente semplici e rigide — che spesso vengono poi applicate in modo aleatorio e incoerente. Peraltro, spesso le classificazioni sono operate in modo assai meno rigoroso**:** alcune classi sono illustrate a fondo, altre meno**;** i loro confini, e persino il loro numero, sono talvolta lasciati vaghi.

L'approccio intensionale è il frutto di un intervento manuale su un qualche settore della realtà e, pertanto, le ontologie per la classificazione vanno intese come strutture funzionali all'organizzazione di oggetti; tuttavia il carattere arbitrario della modellazione delle categorie non implica (o meglio, non dovrebbe implicare) affatto una pretesa che la realtà sia organizzata proprio in quel modo. Questo non comporta affatto la conclusione che ogni ontologia sia equivalente e intercambiabile con ogni altro**;** il confronto va però condotto in base ad criteri eterogenei, che tengono in considerazione l'utilità per specifici scopi cognitivi e/o operativi piuttosto che una valutazione di quanto bene siano bilanciate la sensibilità (che induce a moltiplicare le classi per accrescere la precisione della classificazione) e la parsimonia (che invece suggerisce di limitare il numero delle classi per mantenerne una migliore copertura).

## 4.2 Definizione di criteri di classificazione

La definizione dei criteri di classificazione è un'attività molto complicata perché presuppone la capacità di modellare in che modo la conoscenza esplicita espressa nei documenti sia associata alla conoscenza implicita costituita dalle categorie di interesse. L'esecuzione manuale di tale attività consente il raggiungimento di buoni risultati soprattutto quando il dominio è sufficientemente ristretto ed il linguaggio è caratterizzato da termini tecnici, che non soffrono di problemi di ambiguità. Nell'ultimo decennio, al fine di ottenere prestazioni elevate sono state sviluppate tecniche automatiche per la scoperta di criteri di classificazione (vedi Del. RI.1.02), tipicamente basate su approcci statistici o di ottimizzazione di opportune funzioni obiettivo a partire da un *assegnamento manuale*, ovvero dall'individuazione, per ogni documento di un dato insieme, della opportuna classe di appartenenza.

Nell'ambito della presente metodologia la definizione delle regole di classificazione avviene in maniera semi-automatica, sulla base di due macro-fasi:

- **produzione automatica** di una serie di regole "di default" che esprimono i seguenti meccanismi di classificazione:

- assegna un documento alla categoria *C* se il documento contiene l'nGramma *C* (regola *match*);
    - data due categorie $C_P$ e $C_F$ tali che $C_P$ sia una generalizzazione di $C_F$ (ovvero che $C_F$ sia una specializzazione di $C_P$), assegna a $C_P$ tutti i documenti assegnati a $C_F$ (regola *padre-figlio*);
- **raffinamento manuale** delle regole. In particolare è possibile seguire un approccio basato sulla individuazione di nGrammi *positivi*, cioè la cui presenza nel documento ne caratterizza l'appartenenza alla categoria, e di nGrammi *negativi*, la cui presenza nel documento è indice di non appartenenza alla categoria.

## 4.3 Strumenti software a supporto della metodologia

Per supportare la metodologia sono stati realizzati i due seguenti *acceleratori* software:

- Allo scopo di agevolare la definizione dell'ontologia è stato realizzato il sistema *Ontology Designer*. Questo *tool* si presenta come un'utility a disposizione del knowledge engineer e mette a disposizione le seguenti funzionalità:
    - Definizione della tipologia, applicando tecniche di *clustering* (in particolare l'algoritmo k-means) su un corpus documentale;
    - Definizione manuale dell'ontologia a partire dalla tipologia;
    - Esportazione del'ontologia in diversi formati, al fine di garantire la successiva importazione nel sistema SDCM;
- Per consentire la definizione delle regole di classificazione è stata realizzata l'interfaccia *Rule Editor* integrata nell'applicazione web del sistema SDCM. Tale interfaccia consente:
    - La produzione delle regole di *default* di tipo *match*;
    - La produzione delle regole di *default* di tipo *padre-figlio*;
    - La definizione di nGrammi *positivi* e *negativi* e la generazione automatica di regole che impongono l'assegnazione di un documento alla categoria in presenza di *almeno un* nGramma positivo e *nessun* nGramma negativo;
    - La definizione manuale di regole secondo la sintassi del linguaggio logico Datalog[F] definito nel deliverable RI.1.01.

# 5 Applicazione della metodologia ad un caso d'uso

## 5.1 Dominio di riferimento modello di classificazione

L'area di interesse è quella del pubblico impiego relativamente al servizio del personale di una PAL.

Si è individuato il corpus documentale di riferimento a partire da un corpus di 16500 delibere del Comune di Paola (CS) da cui è stato estratto un campione di 778 documenti per costruire il corpus di test.

## 5.2 Definizione Schema di classificazione

### 5.2.1 *Tipologia*

La tipologia è stata ottenuta utilizzando il sistema *Ontology Designer* descritto in precedenza con riferimento ai 778 documenti costituenti il corpus di test. La figura seguente mostra le modalità di configurazione del sistema.

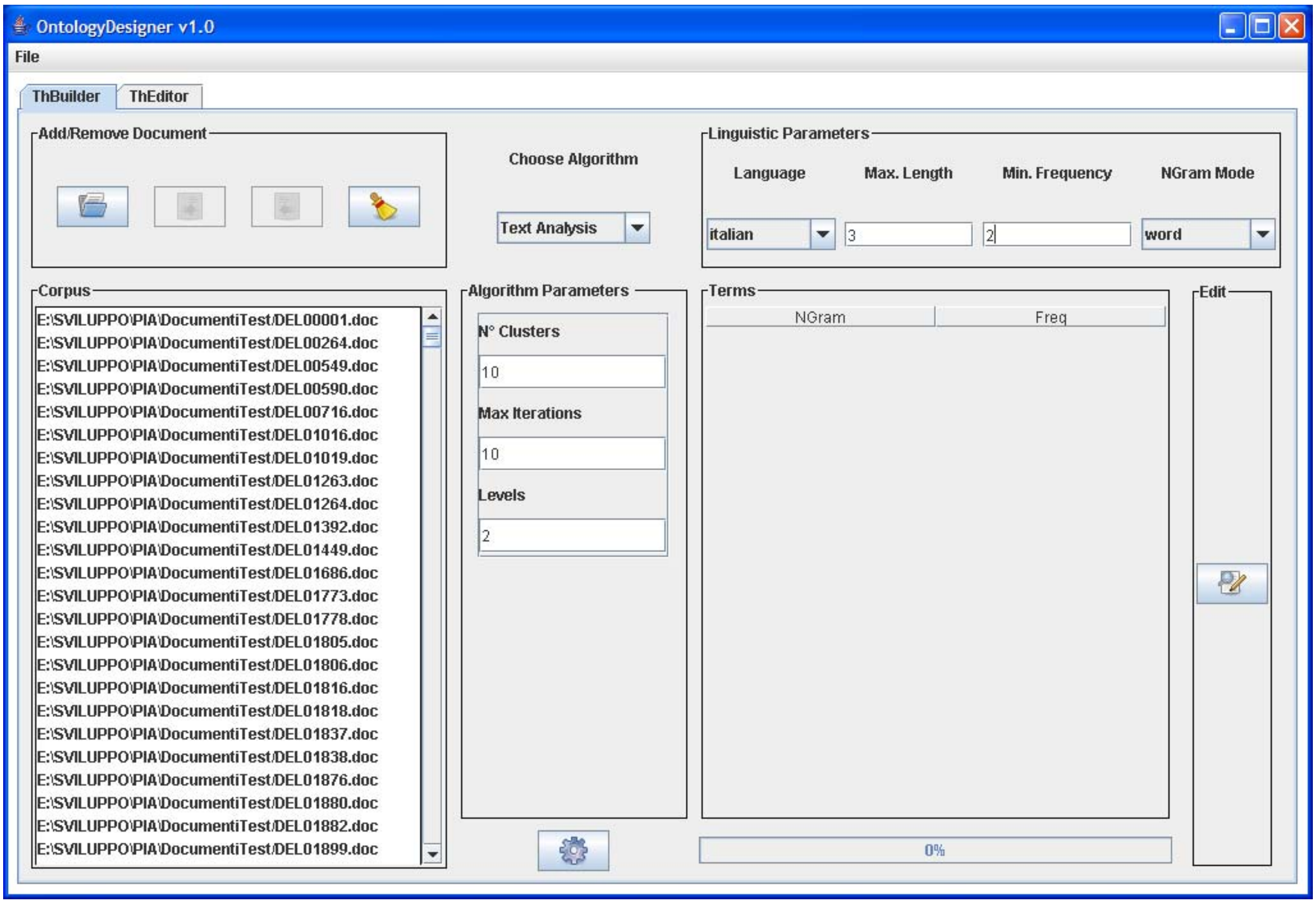

Le figure seguenti mostrano la tipologia ottenuta. Per ogni classe il sistema *Ontology Designer* restituisce inoltre i 15 termini più rappresentativi, attraverso i quali è possibile dedurre i contenuti dei documenti inseriti nella classe stessa.

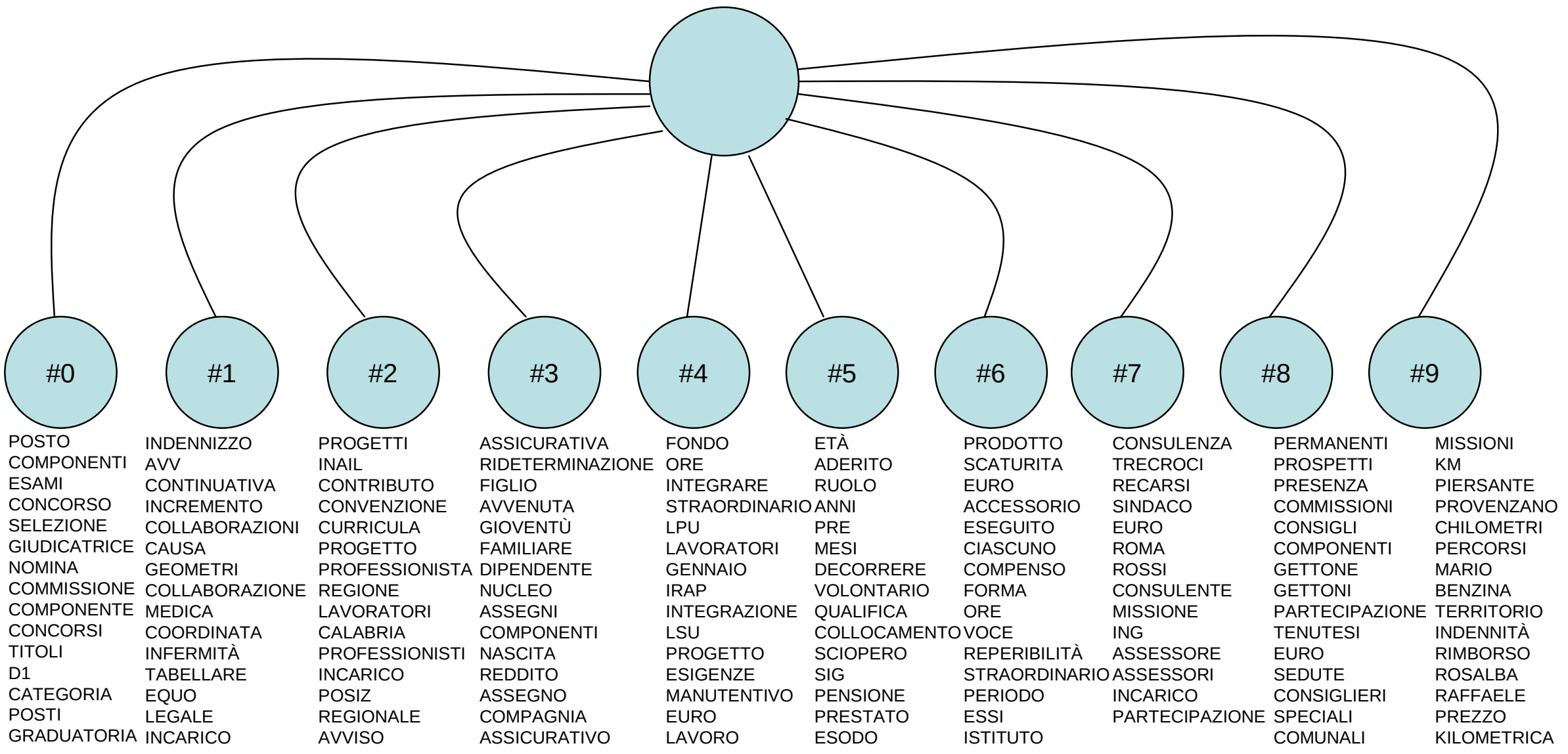

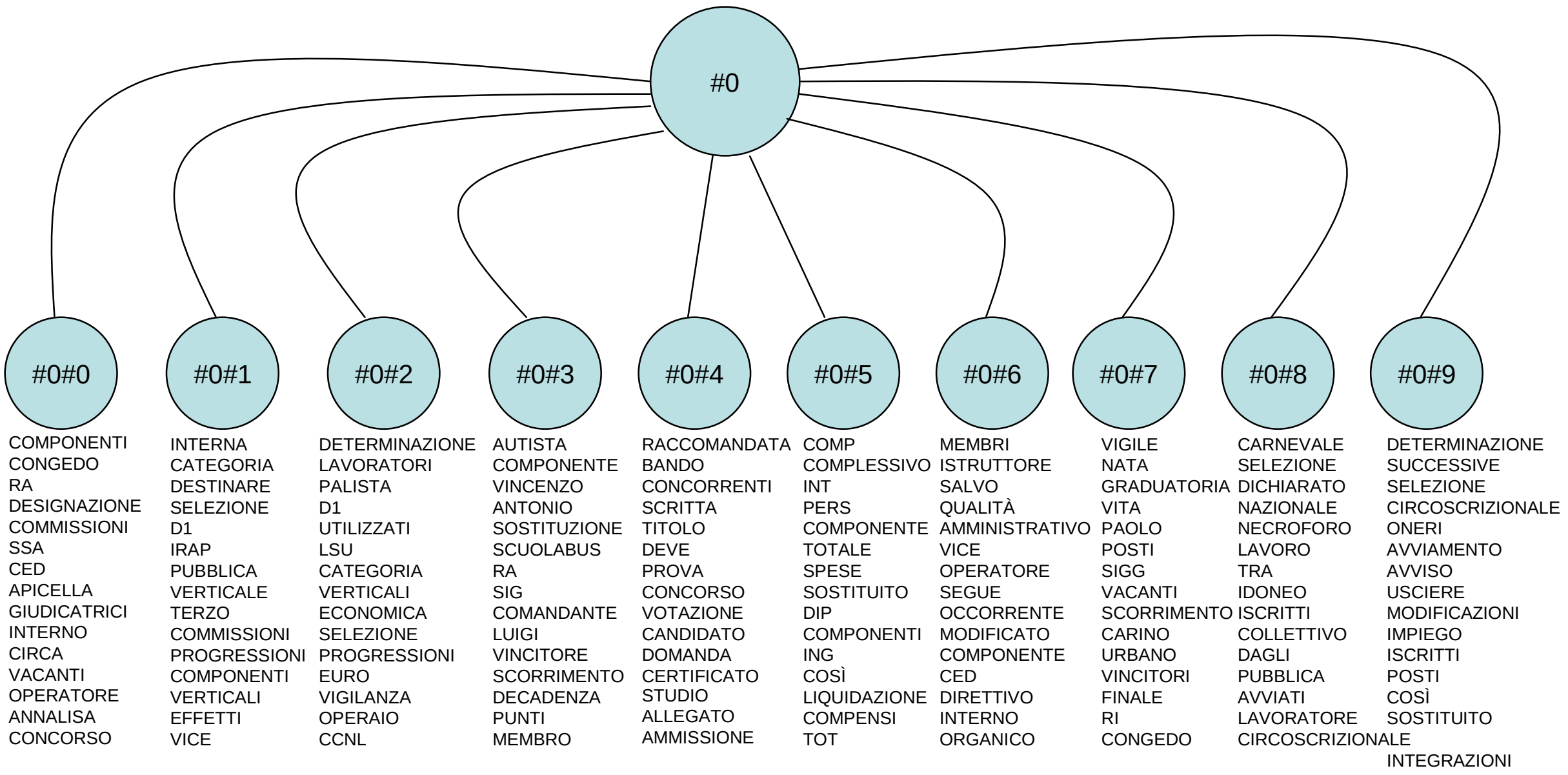

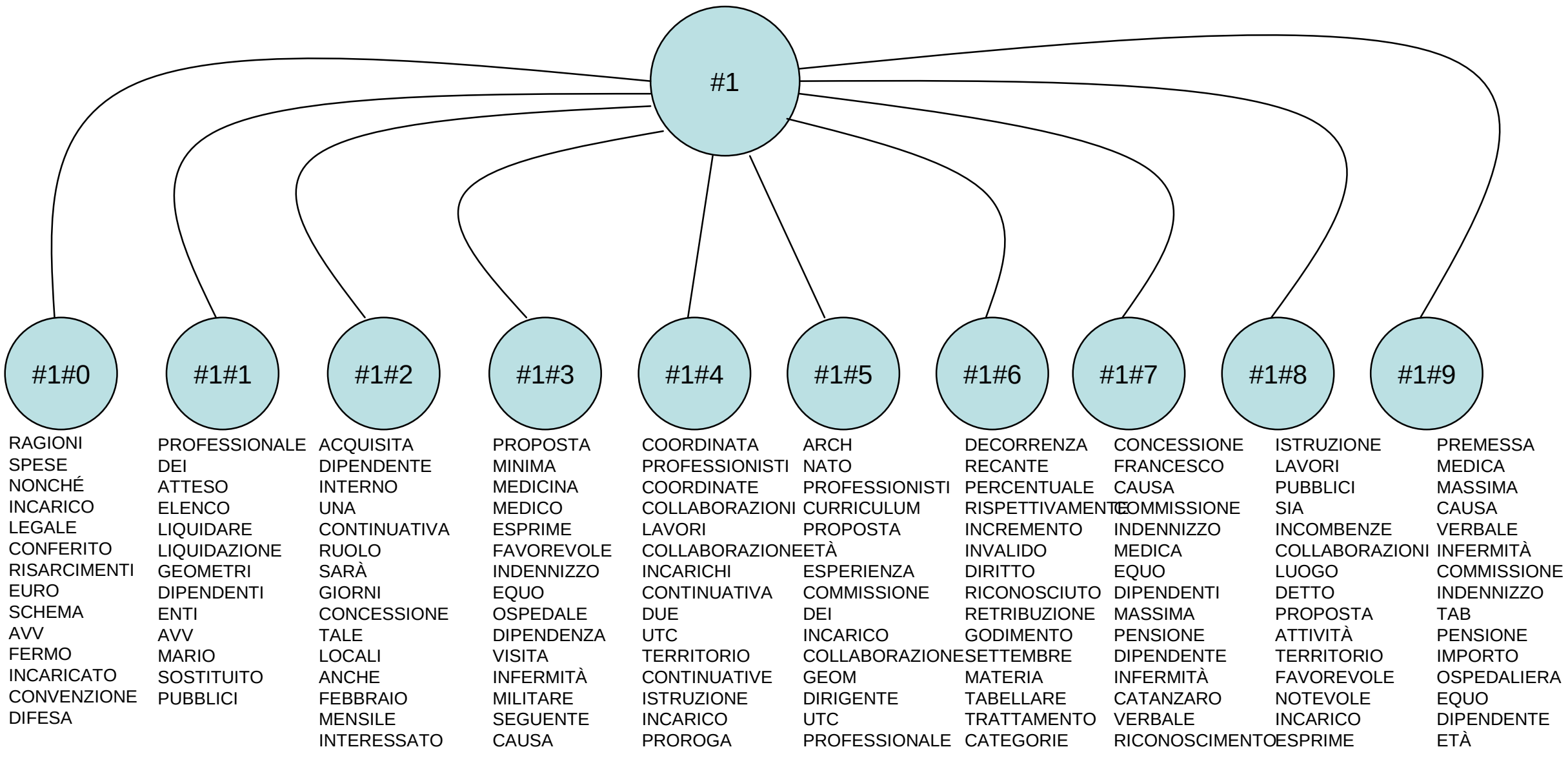

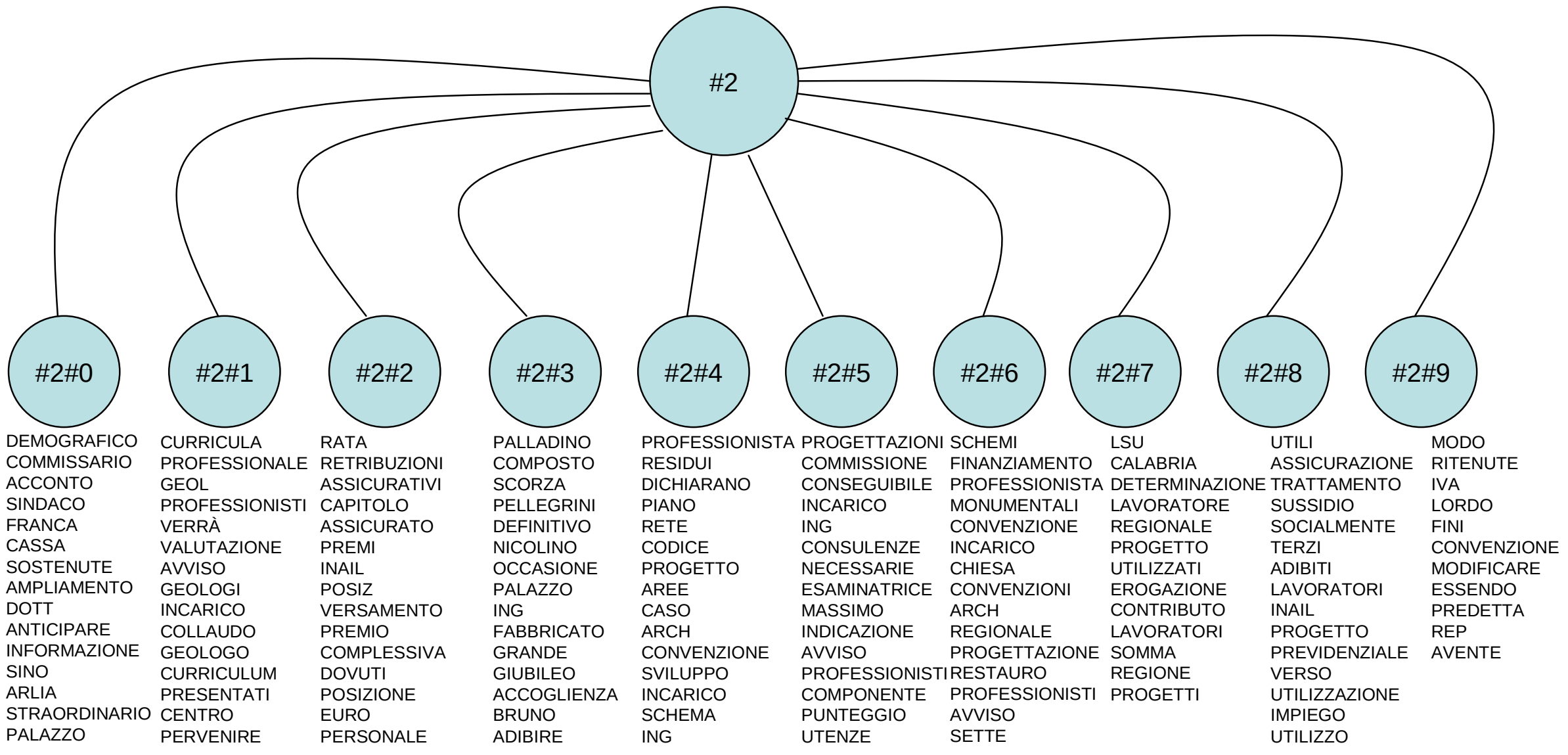

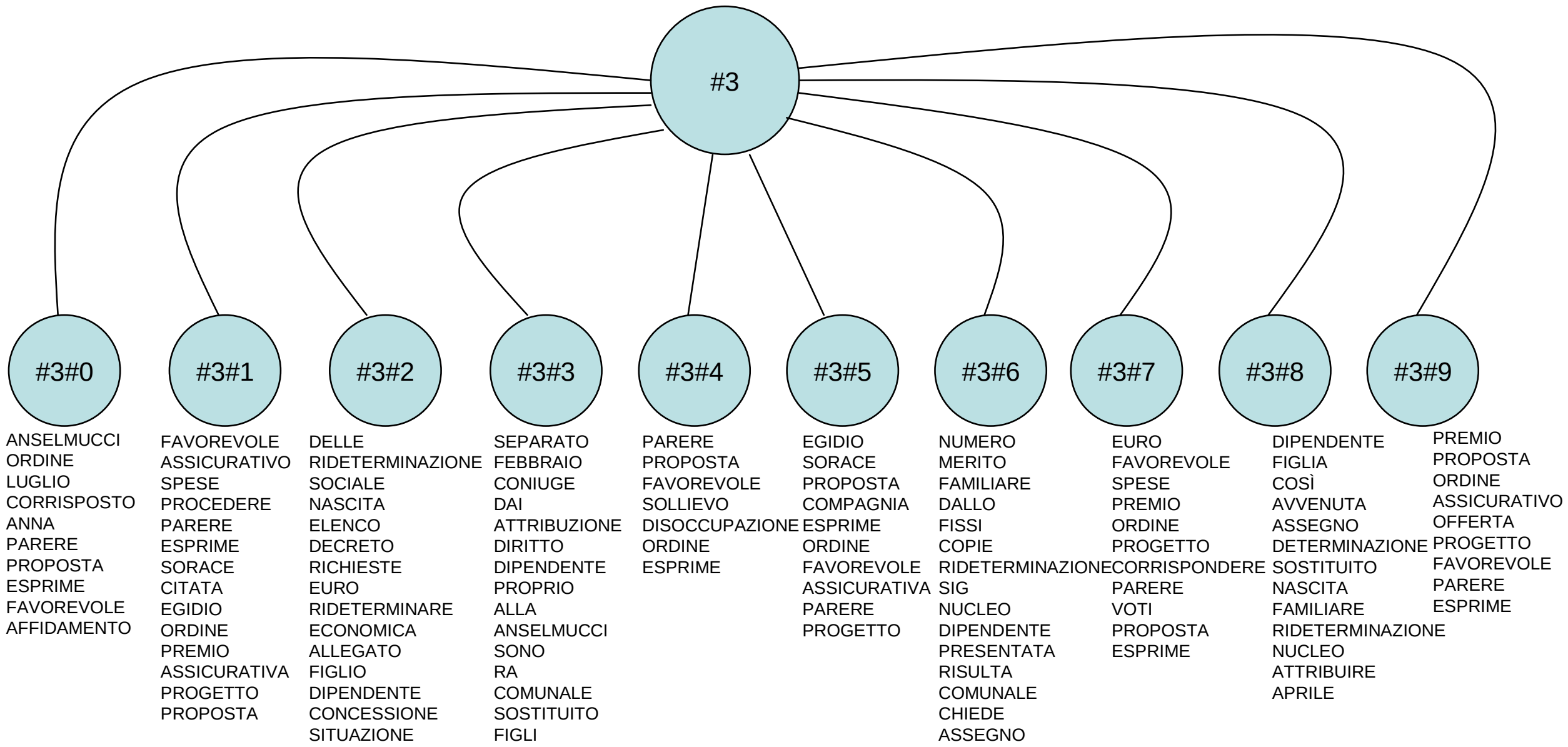
#3
#3#0
#3#1
#3#2
#3#3
#3#4
#3#5
#3#6
#3#7
#3#8
#3#9
ANSELMUCCI ORDINE LUGLIO CORRISPOSTO ANNA PARERE PROPOSTA ESPRIME FAVOREVOLE AFFIDAMENTO
FAVOREVOLE ASSICURATIVO SPESE PROCEDERE PARERE ESPRIME SORACE CITATA EGIDIO ORDINE PREMIO ASSICURATIVA PROGETTO PROPOSTA
DELLE RIDETERMINAZIONE SOCIALE NASCITA ELENCO DECRETO RICHIESTE EURO RIDETERMINARE ECONOMICA ALLEGATO FIGLIO DIPENDENTE CONCESSIONE SITUAZIONE
SEPARATO FEBBRAIO CONIUGE DAI ATTRIBUZIONE DIRITTO DIPENDENTE PROPRIO ALLA ANSELMUCCI SONO RA COMUNALE SOSTITUITO FIGLI
PARERE PROPOSTA FAVOREVOLE SOLLIEVO DISOCCUPAZIONE ORDINE ESPRIME
EGIDIO SORACE PROPOSTA COMPAGNIA ESPRIME ORDINE FAVOREVOLE ASSICURATIVA PARERE PROGETTO
NUMERO MERITO FAMILIARE DALLO FISSI COPIE RIDETERMINAZIONE SIG NUCLEO DIPENDENTE PRESENTATA RISULTA COMUNALE CHIEDE ASSEGNO
EURO FAVOREVOLE SPESE PREMIO ORDINE PROGETTO CORRISPONDERE PARERE VOTI PROPOSTA ESPRIME
DIPENDENTE FIGLIA COSÌ AVVENUTA ASSEGNO DETERMINAZIONE SOSTITUITO NASCITA FAMILIARE RIDETERMINAZIONE NUCLEO ATTRIBUIRE APRILE
PREMIO PROPOSTA ORDINE ASSICURATIVO OFFERTA PROGETTO FAVOREVOLE PARERE ESPRIME

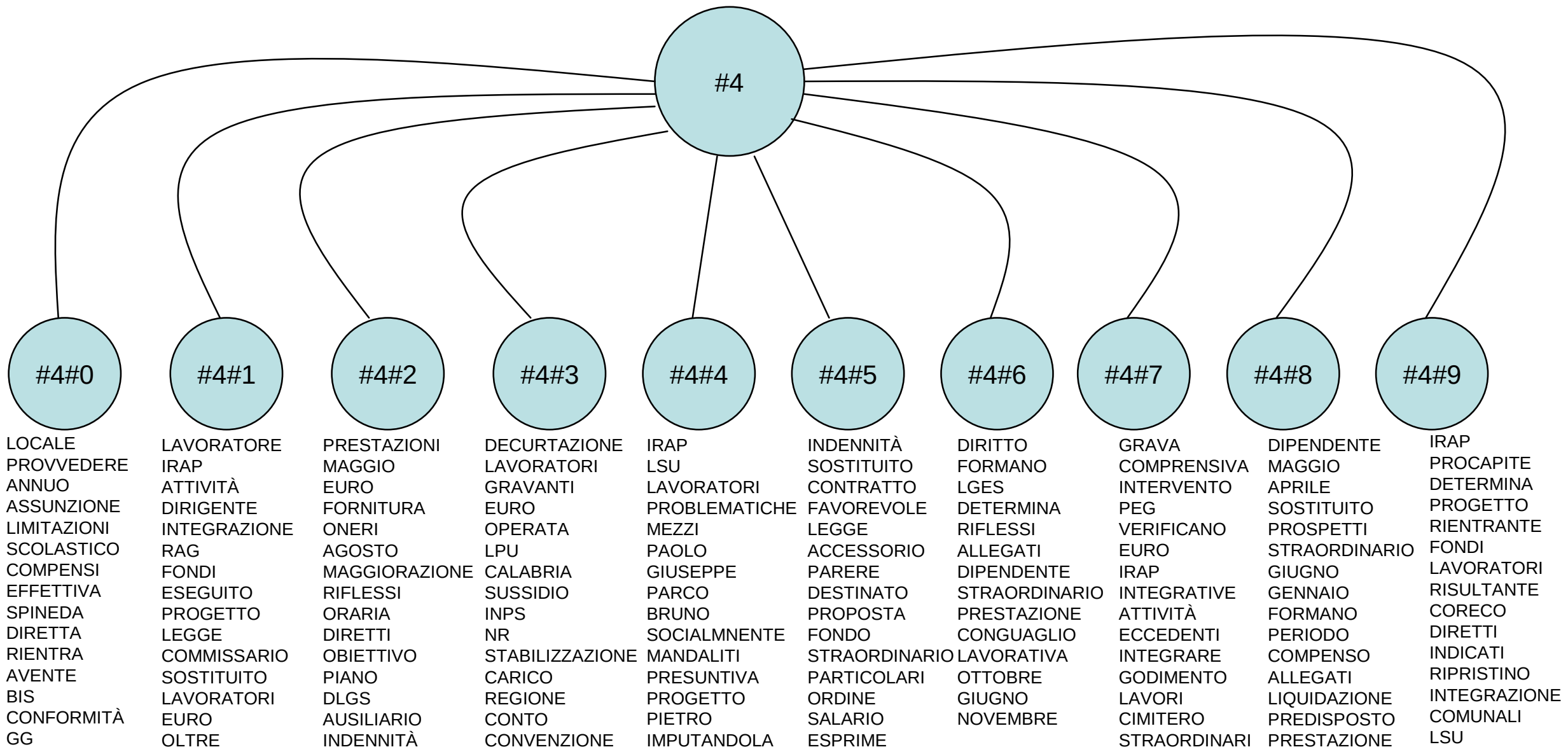
#4
#4#0
#4#1
#4#2
#4#3
#4#4
#4#5
#4#6
#4#7
#4#8
#4#9
LOCALE PROVVEDERE ANNUO ASSUNZIONE LIMITAZIONI SCOLASTICO COMPENSI EFFETTIVA SPINEDA DIRETTA RIENTRA AVENTE BIS CONFORMITÀ GG
LAVORATORE IRAP ATTIVITÀ DIRIGENTE INTEGRAZIONE RAG FONDI ESEGUITO PROGETTO LEGGE COMMISSARIO SOSTITUITO LAVORATORI EURO OLTRE
PRESTAZIONI MAGGIO EURO FORNITURA ONERI AGOSTO MAGGIORAZIONE RIFLESSI ORARIA DIRETTI OBIETTIVO PIANO DLGS AUSILIARIO INDENNITÀ
DECURTAZIONE LAVORATORI GRAVANTI EURO OPERATA LPU CALABRIA SUSSIDIO INPS NR STABILIZZAZIONE CARICO REGIONE CONTO CONVENZIONE
IRAP LSU LAVORATORI PROBLEMATICHE MEZZI PAOLO GIUSEPPE PARCO BRUNO SOCIALMNENTE MANDALITI PRESUNTIVA PROGETTO PIETRO IMPUTANDOLA
INDENNITÀ SOSTITUITO CONTRATTO FAVOREVOLE LEGGE ACCESSORIO PARERE DESTINATO PROPOSTA FONDO STRAORDINARIO PARTICOLARI ORDINE SALARIO ESPRIME
DIRITTO FORMANO LGES DETERMINA RIFLESSI ALLEGATI DIPENDENTE STRAORDINARIO PRESTAZIONE CONGUAGLIO LAVORATIVA OTTOBRE GIUGNO NOVEMBRE
GRAVA COMPRENSIVA INTERVENTO PEG VERIFICANO EURO IRAP INTEGRATIVE ATTIVITÀ ECCEDENTI INTEGRARE GODIMENTO LAVORI CIMITERO STRAORDINARI
DIPENDENTE MAGGIO APRILE SOSTITUITO PROSPETTI STRAORDINARIO GIUGNO GENNAIO FORMANO PERIODO COMPENSO ALLEGATI LIQUIDAZIONE PREDISPOSTO PRESTAZIONE
IRAP PROCAPITE DETERMINA PROGETTO RIENTRANTE FONDI LAVORATORI RISULTANTE CORECO DIRETTI INDICATI RIPRISTINO INTEGRAZIONE COMUNALI LSU

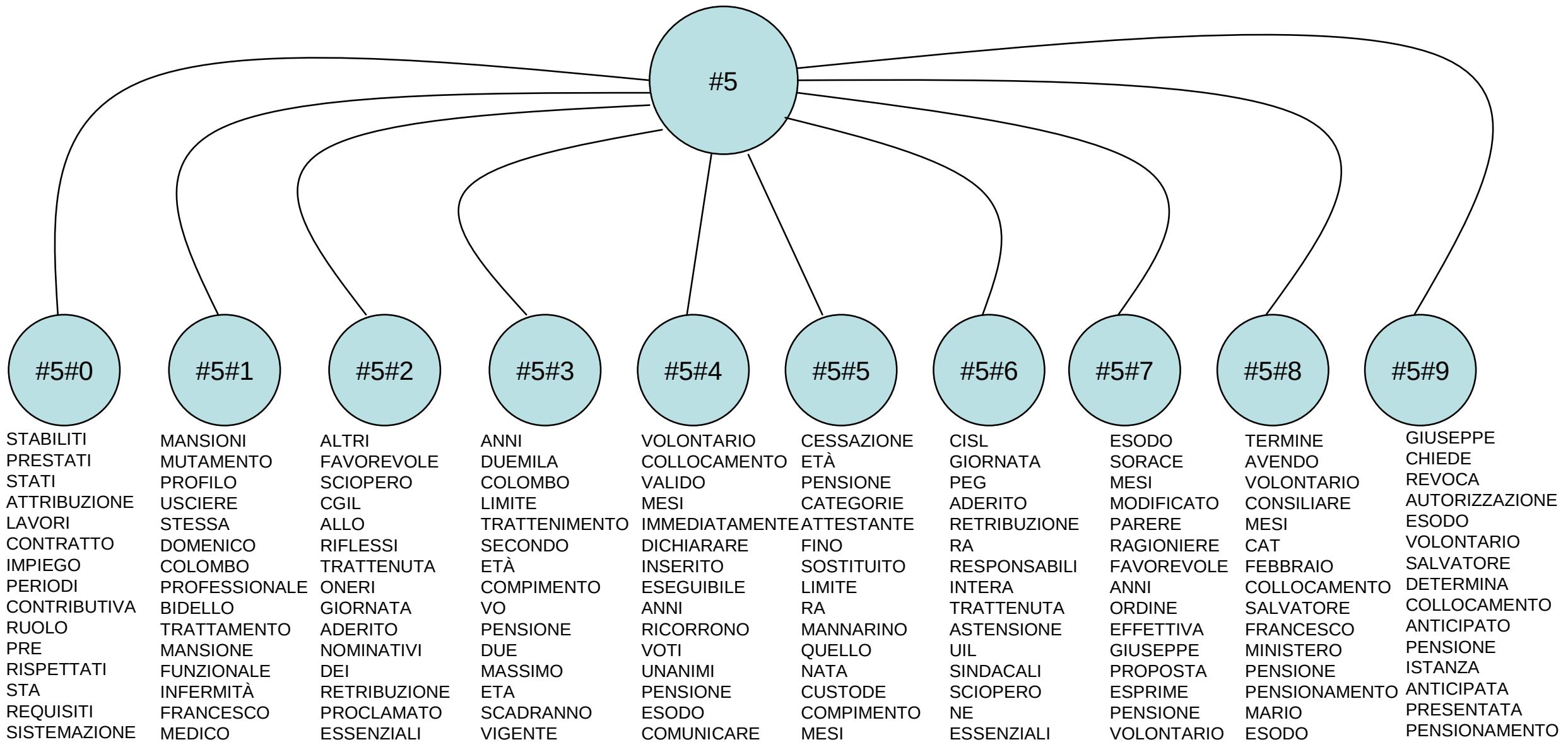
#5
#5#0
STABILITI PRESTATI STATI ATTRIBUZIONE LAVORI CONTRATTO IMPIEGO PERIODI CONTRIBUTIVA RUOLO PRE RISPETTATI STA REQUISITI SISTEMAZIONE
#5#1
MANSIONI MUTAMENTO PROFILO USCIERE STESSA DOMENICO COLOMBO PROFESSIONALE BIDELLO TRATTAMENTO MANSIONE FUNZIONALE INFERMITÀ FRANCESCO MEDICO
#5#2
ALTRI FAVOREVOLE SCIOPERO CGIL ALLO RIFLESSI TRATTENUTA ONERI GIORNATA ADERITO NOMINATIVI DEI RETRIBUZIONE PROCLAMATO ESSENZIALI
#5#3
ANNI DUEMILA COLOMBO LIMITE TRATTENIMENTO SECONDO ETÀ COMPIMENTO VO PENSIONE DUE MASSIMO ETA SCADRANNO VIGENTE
#5#4
VOLONTARIO COLLOCAMENTO VALIDO MESI IMMEDIATAMENTE DICHIARARE INSERITO ESEGUIBILE ANNI RICORRONO VOTI UNANIMI PENSIONE ESODO COMUNICARE
#5#5
CESSAZIONE ETÀ PENSIONE CATEGORIE ATTESTANTE FINO SOSTITUITO LIMITE RA MANNARINO QUELLO NATA CUSTODE COMPIMENTO MESI
#5#6
CISL GIORNATA PEG ADERITO RETRIBUZIONE RA RESPONSABILI INTERA TRATTENUTA ASTENSIONE UIL SINDACALI SCIOPERO NE ESSENZIALI
#5#7
ESODO SORACE MESI MODIFICATO PARERE RAGIONIERE FAVOREVOLE ANNI ORDINE EFFETTIVA GIUSEPPE PROPOSTA ESPRIME PENSIONE VOLONTARIO
#5#8
TERMINE AVENDO VOLONTARIO CONSILIARE MESI CAT FEBBRAIO COLLOCAMENTO SALVATORE FRANCESCO MINISTERO PENSIONE PENSIONAMENTO MARIO ESODO
#5#9
GIUSEPPE CHIEDE REVOCA AUTORIZZAZIONE ESODO VOLONTARIO SALVATORE DETERMINA COLLOCAMENTO ANTICIPATO PENSIONE ISTANZA ANTICIPATA PRESENTATA PENSIONAMENTO

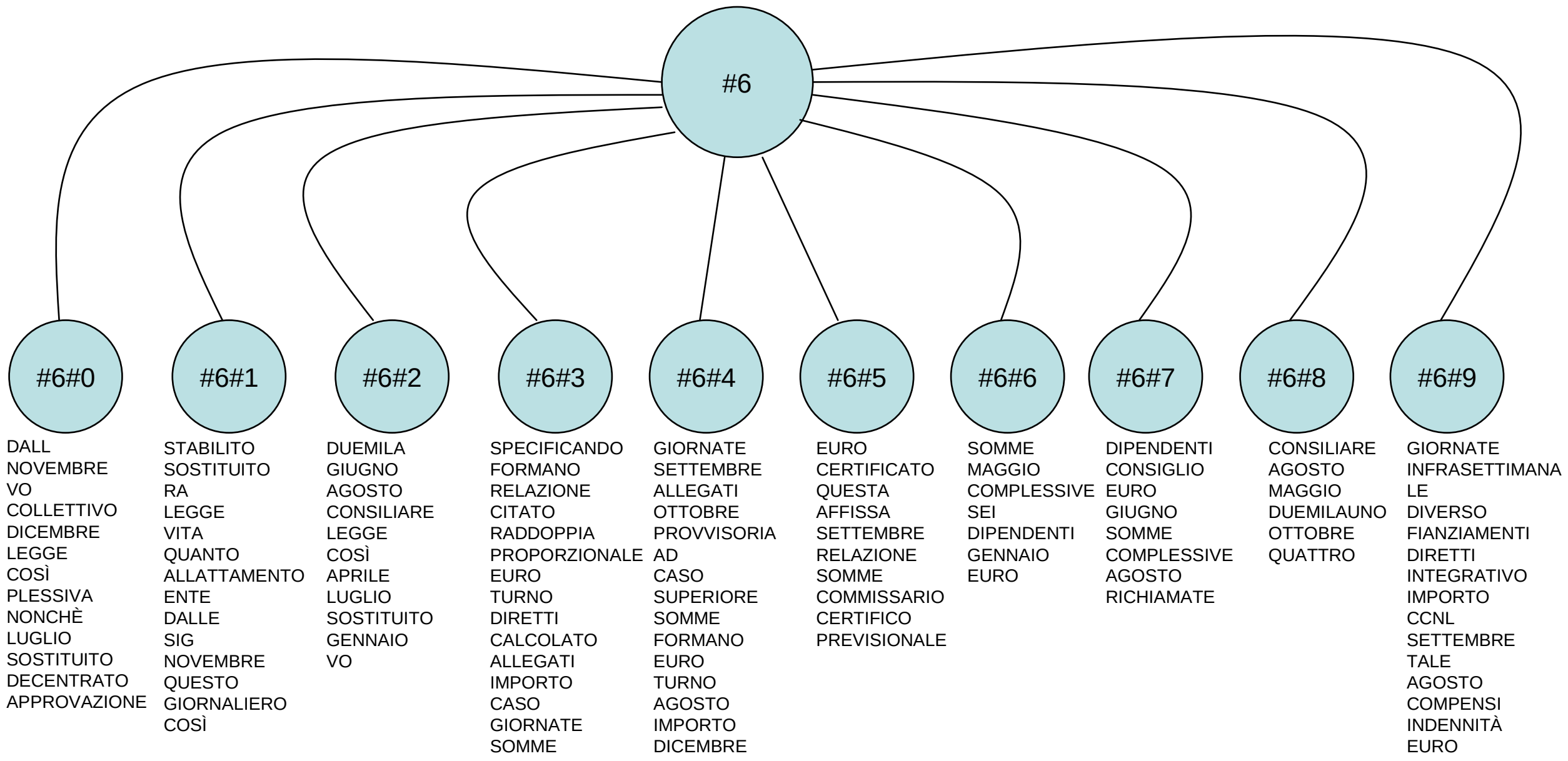
#6
#6#0
DALL NOVEMBRE VO COLLETTIVO DICEMBRE LEGGE COSÌ PLESSIVA NONCHÈ LUGLIO SOSTITUITO DECENTRATO APPROVAZIONE
#6#1
STABILITO SOSTITUITO RA LEGGE VITA QUANTO ALLATTAMENTO ENTE DALLE SIG NOVEMBRE QUESTO GIORNALIERO COSÌ
#6#2
DUEMILA GIUGNO AGOSTO CONSILIARE LEGGE COSÌ APRILE LUGLIO SOSTITUITO GENNAIO VO
#6#3
SPECIFICANDO FORMANO RELAZIONE CITATO RADDOPPIA PROPORZIONALE EURO TURNO DIRETTI CALCOLATO ALLEGATI IMPORTO CASO GIORNATE SOMME
#6#4
GIORNATE SETTEMBRE ALLEGATI OTTOBRE PROVVISORIA AD CASO SUPERIORE SOMME FORMANO EURO TURNO AGOSTO IMPORTO DICEMBRE
#6#5
EURO CERTIFICATO QUESTA AFFISSA SETTEMBRE RELAZIONE SOMME COMMISSARIO CERTIFICO PREVISIONALE
#6#6
SOMME MAGGIO COMPLESSIVE SEI DIPENDENTI GENNAIO EURO
#6#7
DIPENDENTI CONSIGLIO EURO GIUGNO SOMME COMPLESSIVE AGOSTO RICHIAMATE
#6#8
CONSILIARE AGOSTO MAGGIO DUEMILAUNO OTTOBRE QUATTRO
#6#9
GIORNATE INFRASETTIMANA LE DIVERSO FIANZIAMENTI DIRETTI INTEGRATIVO IMPORTO CCNL SETTEMBRE TALE AGOSTO COMPENSI INDENNITÀ EURO

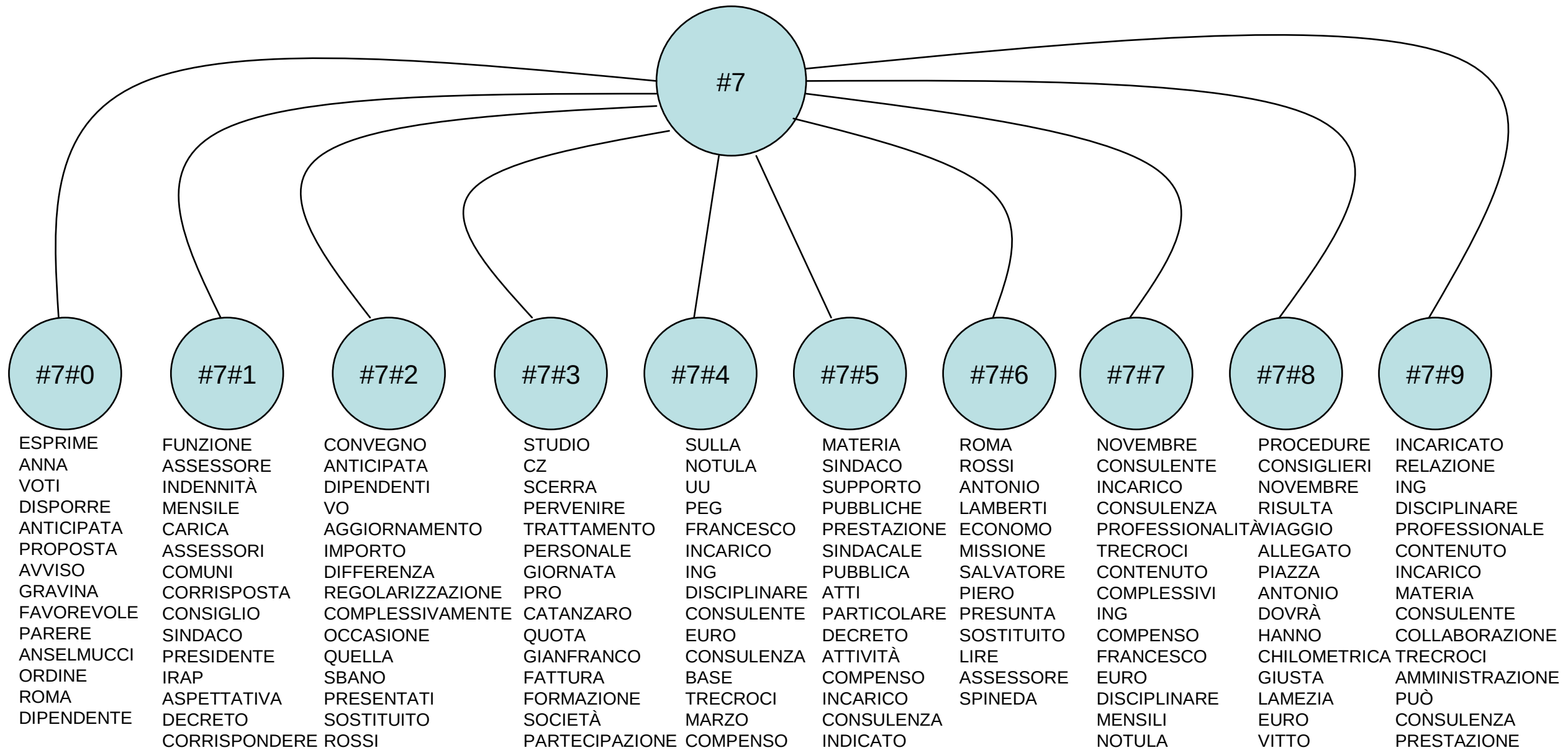
#7
#7#0
ESPRIME ANNA VOTI DISPORRE ANTICIPATA PROPOSTA AVVISO GRAVINA FAVOREVOLE PARERE ANSELMUCCI ORDINE ROMA DIPENDENTE
#7#1
FUNZIONE ASSESSORE INDENNITÀ MENSILE CARICA ASSESSORI COMUNI CORRISPOSTA CONSIGLIO SINDACO PRESIDENTE IRAP ASPETTATIVA DECRETO CORRISPONDERE
#7#2
CONVEGNO ANTICIPATA DIPENDENTI VO AGGIORNAMENTO IMPORTO DIFFERENZA REGOLARIZZAZIONE COMPLESSIVAMENTE OCCASIONE QUELLA SBANO PRESENTATI SOSTITUITO ROSSI
#7#3
STUDIO CZ SCERRA PERVENIRE TRATTAMENTO PERSONALE GIORNATA PRO CATANZARO QUOTA GIANFRANCO FATTURA FORMAZIONE SOCIETÀ PARTECIPAZIONE
#7#4
SULLA NOTULA UU PEG FRANCESCO INCARICO ING DISCIPLINARE CONSULENTE EURO CONSULENZA BASE TRECROCI MARZO COMPENSO
#7#5
MATERIA SINDACO SUPPORTO PUBBLICHE PRESTAZIONE SINDACALE PUBBLICA ATTI PARTICOLARE DECRETO ATTIVITÀ COMPENSO INCARICO CONSULENZA INDICATO
#7#6
ROMA ROSSI ANTONIO LAMBERTI ECONOMO MISSIONE SALVATORE PIERO PRESUNTA SOSTITUITO LIRE ASSESSORE SPINEDA
#7#7
NOVEMBRE CONSULENTE INCARICO CONSULENZA PROFESSIONALITÀ TRECROCI CONTENUTO COMPLESSIVI ING COMPENSO FRANCESCO EURO DISCIPLINARE MENSILI NOTULA
#7#8
PROCEDURE CONSIGLIERI NOVEMBRE RISULTA VIAGGIO ALLEGATO PIAZZA ANTONIO DOVRÀ HANNO CHILOMETRICA GIUSTA LAMEZIA EURO VITTO
#7#9
INCARICATO RELAZIONE ING DISCIPLINARE PROFESSIONALE CONTENUTO INCARICO MATERIA CONSULENTE COLLABORAZIONE TRECROCI AMMINISTRAZIONE PUÒ CONSULENZA PRESTAZIONE

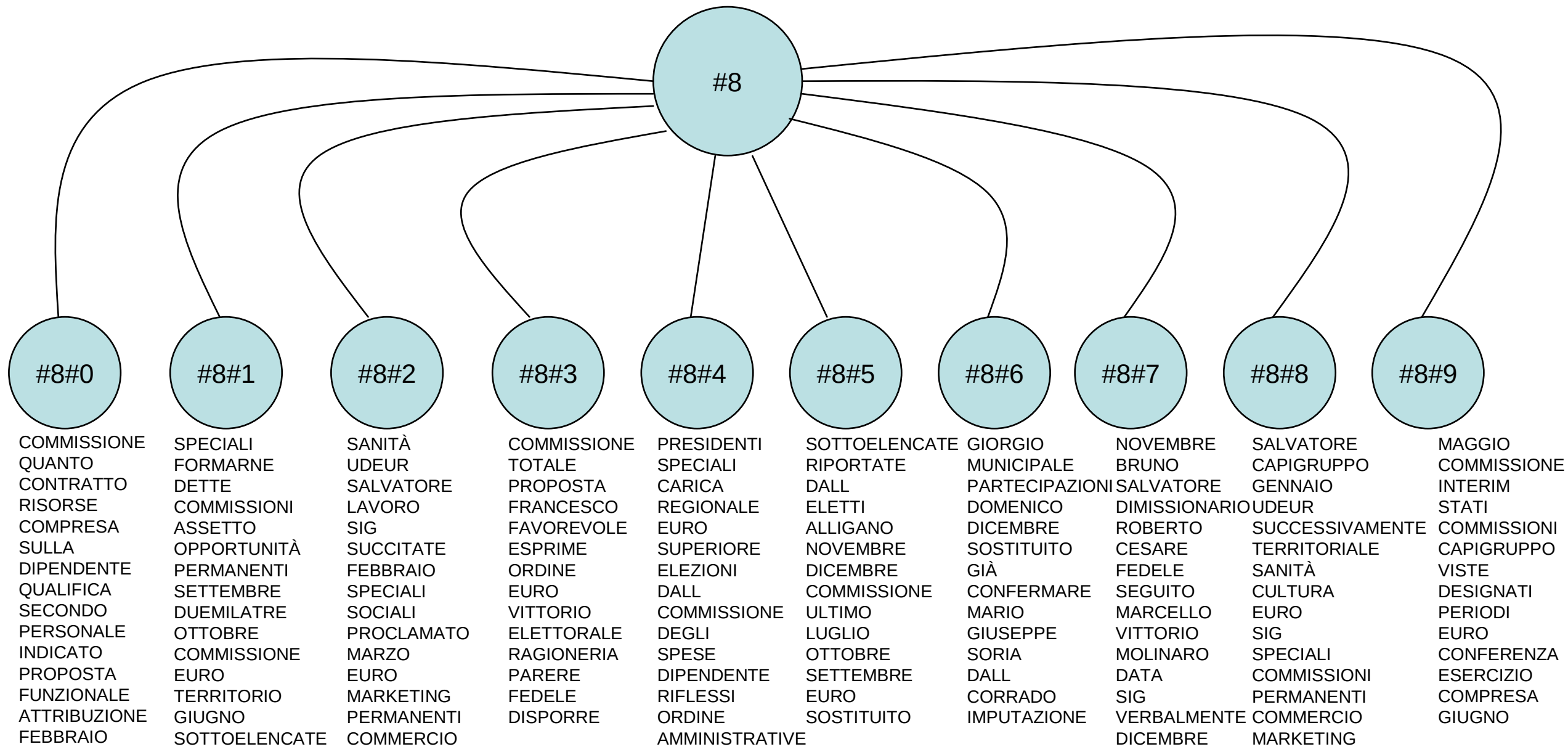
#8
#8#0
COMMISSIONE QUANTO CONTRATTO RISORSE COMPRESA SULLA DIPENDENTE QUALIFICA SECONDO PERSONALE INDICATO PROPOSTA FUNZIONALE ATTRIBUZIONE FEBBRAIO
#8#1
SPECIALI FORMARNE DETTE COMMISSIONI ASSETTO OPPORTUNITÀ PERMANENTI SETTEMBRE DUEMILATRE OTTOBRE COMMISSIONE EURO TERRITORIO GIUGNO SOTTOELENCATE
#8#2
SANITÀ UDEUR SALVATORE LAVORO SIG SUCCITATE FEBBRAIO SPECIALI SOCIALI PROCLAMATO MARZO EURO MARKETING PERMANENTI COMMERCIO
#8#3
COMMISSIONE TOTALE PROPOSTA FRANCESCO FAVOREVOLE ESPRIME ORDINE EURO VITTORIO ELETTORALE RAGIONERIA PARERE FEDELE DISPORRE
#8#4
PRESIDENTI SPECIALI CARICA REGIONALE EURO SUPERIORE ELEZIONI DALL COMMISSIONE DEGLI SPESE DIPENDENTE RIFLESSI ORDINE AMMINISTRATIVE
#8#5
SOTTOELENCATE RIPORTATE DALL ELETTI ALLIGANO NOVEMBRE DICEMBRE COMMISSIONE ULTIMO LUGLIO OTTOBRE SETTEMBRE EURO SOSTITUITO
#8#6
GIORGIO MUNICIPALE PARTECIPAZIONI DOMENICO DICEMBRE SOSTITUITO GIÀ CONFERMARE MARIO GIUSEPPE SORIA DALL CORRADO IMPUTAZIONE
#8#7
NOVEMBRE BRUNO SALVATORE DIMISSIONARIO ROBERTO CESARE FEDELE SEGUITO MARCELLO VITTORIO MOLINARO DATA SIG VERBALMENTE DICEMBRE
#8#8
SALVATORE CAPIGRUPPO GENNAIO UDEUR SUCCESSIVAMENTE TERRITORIALE SANITÀ CULTURA EURO SIG SPECIALI COMMISSIONI PERMANENTI COMMERCIO MARKETING
#8#9
MAGGIO COMMISSIONE INTERIM STATI COMMISSIONI CAPIGRUPPO VISTE DESIGNATI PERIODI EURO CONFERENZA ESERCIZIO COMPRESA GIUGNO

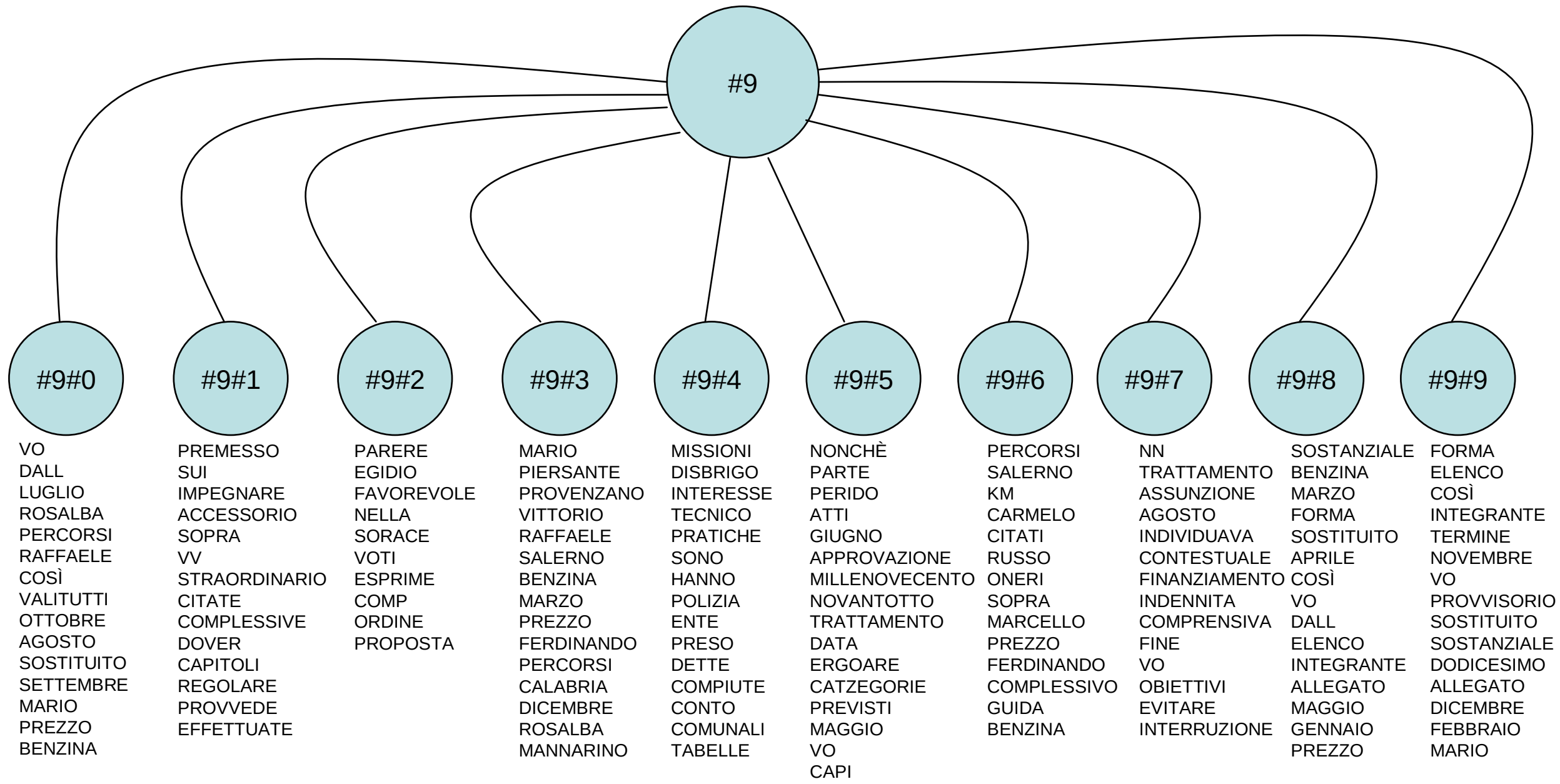

### *5.2.2 Ontologia*

L'ontologia è stata ottenuto a partire dalla tipologia, eseguendo attività di: (i) riduzione della tipologia; (ii) aggregazione; (iii) generalizzazione e (iv) specializzazione.

**Passo 1: Riduzione della tipologia**

L'esame dei termini caratterizzanti consente di individuare le classi della tipologia che contengono documenti sicuramente non attinenti al dominio di riferimento che si intende modellare. Le categorie escluse sono le seguenti: #0#3, #0#6, #0#7, #0#8, #0#9, #1#2, #1#3, #1#5, #1#7, #1#9, #2#0, #2#3, #2#4, #2#6, #2#9, #3 e tutte le categorie figlie, #4#0, #4#2, #4#7, #5#0, #5#1, #6 e tutte le categorie figlie, #7#0, #7#2, #7#3, #8#1, #8#2, #8#3, #8#4, #8#5, #8#6, #8#7, #8#8, #8#9, #9#0, #9#2, #9#3, #9#4, #9#5, #9#7, #9#8, #9#9.

**Passo 2: Aggregazione**

Sulla base della coerenza semantica dei termini caratterizzanti le singole categorie, è possibile definire un insieme di nodi "di sintesi" secondo la seguente tabella:

| Nodo di sintesi | Nodi aggregati |
|---|---|
| #A | #0#0, #0#4 |
| #B | #1#0, #1#1, #1#4, #1#8 |
| #C | #2#7, #2#8 |
| #D | #4#3, #4#4, #4#9 |
| #E | #4#5, #4#6, #4#8 |
| #F | #5#2, #5#6 |
| #G | #5#4, #5#5, #5#7, #5#8, #5#9 |
| #H | #7#4, #7#5, #7#7 |
| #I | #7#6, #7#8 |

La tipologia ridotta si presenta come nella seguente figura.

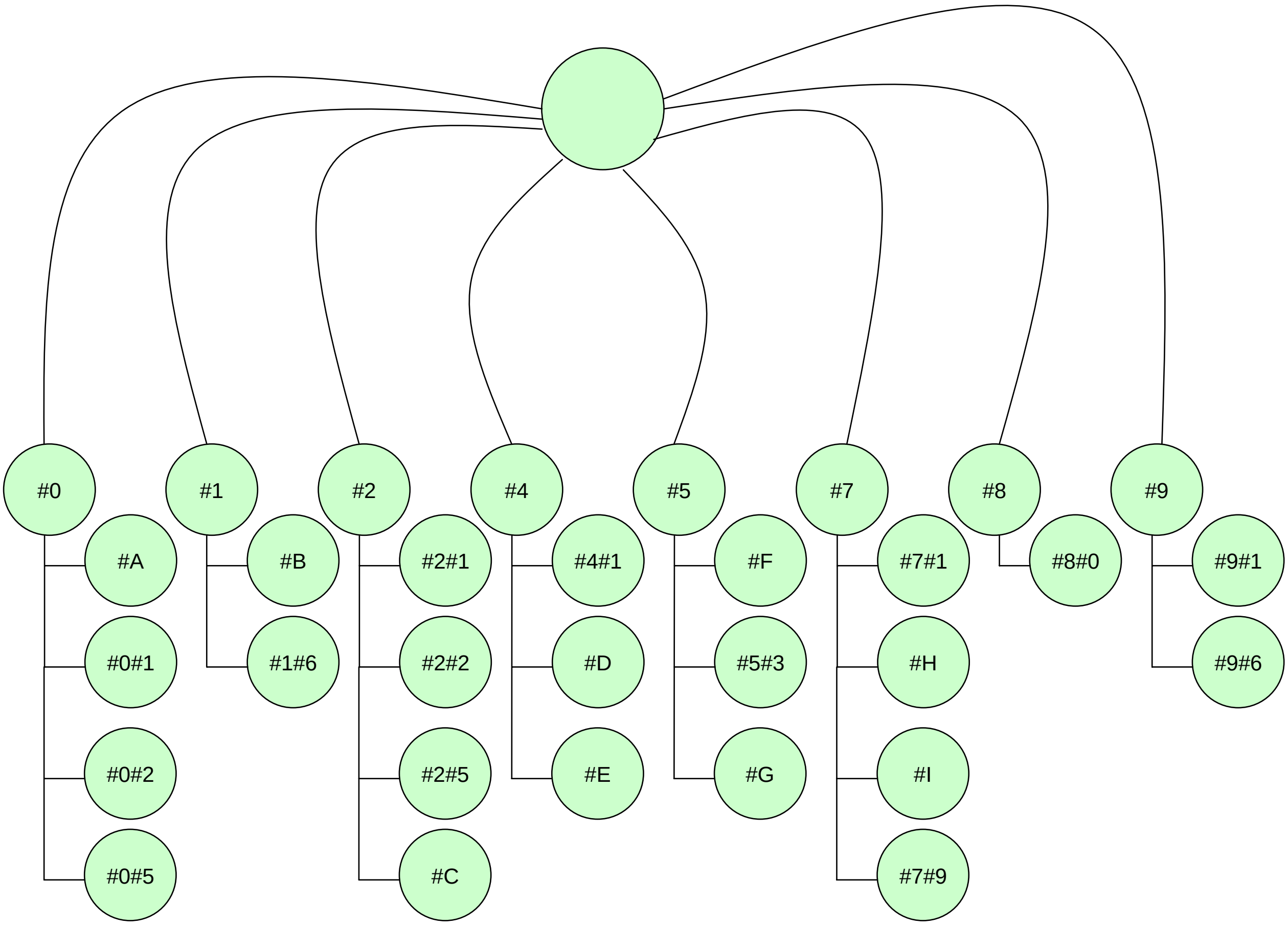

L'operazione di aggregazione può essere iterata ulteriormente tenendo in considerazione che: (i) le classi di primo livello (#0, #1,…) possono essere aggregate poiché hanno un ruolo meramente formale, nel senso che sono calcolate nella fase di clustering in modo da garantire un'iterazione a due livelli; (ii) diverse classi condividono i termini più significativi che caratterizzano i documenti classificati al loro interno. In particolare, per valutare il punto (ii), la tabella seguente riporta i termini più significativi per le classi residue della tipologia. Per ogni riga, è individuato il nome della classe rappresentativa.

| Classe | Termini più rilevanti | Classe |
|---|---|---|
| #A | Concorso | Concorso |
| #0#1 | Categoria | Categorie Personale |
| #0#2 | LSU | LSU/LPU |
| #0#5 | Compensi, commissione | Compensi a commissioni |
| #B | Spese, liquidare, risarcimenti | Rimborso spese |
| #1#6 | Retribuzione | Retribuzione |
| #2#1 | Incarico, professionisti, curricula | Consulenti |
| #2#2 | Retribuzioni | Retribuzione |
| #2#5 | Commissione, esaminatrice | Concorso |
| #C | LSU | LSU/LPU |
| #4#1 | Dirigente | Categorie Personale |
| #D | LSU, LPU | LSU/LPU |
| #E | Straordinario | Straordinari |
| #F | Sciopero, sindacati | Sciopero |
| #5#3 | Trattenimento, pensione, limite | Trattenimento in servizio |
| #G | Pensione, esodo | Esodo |
| #7#1 | Indennità, assessori | Compensi ad amministratori |
| #H | Consulenza, compenso | Compensi a liberi professionisti |
| #I | Missione, chilometrico | Rimborso spese |
| #7#9 | Consulente, collaborazione, prestazione | Consulenti |
| #8#0 | Commissione | Concorso |
| #9#1 | Straordinario | Straordinari |
| #9#6 | KM, benzina | Rimborso spese |

La tipologia dopo quest’ultima operazione di riduzione diventa la seguente.

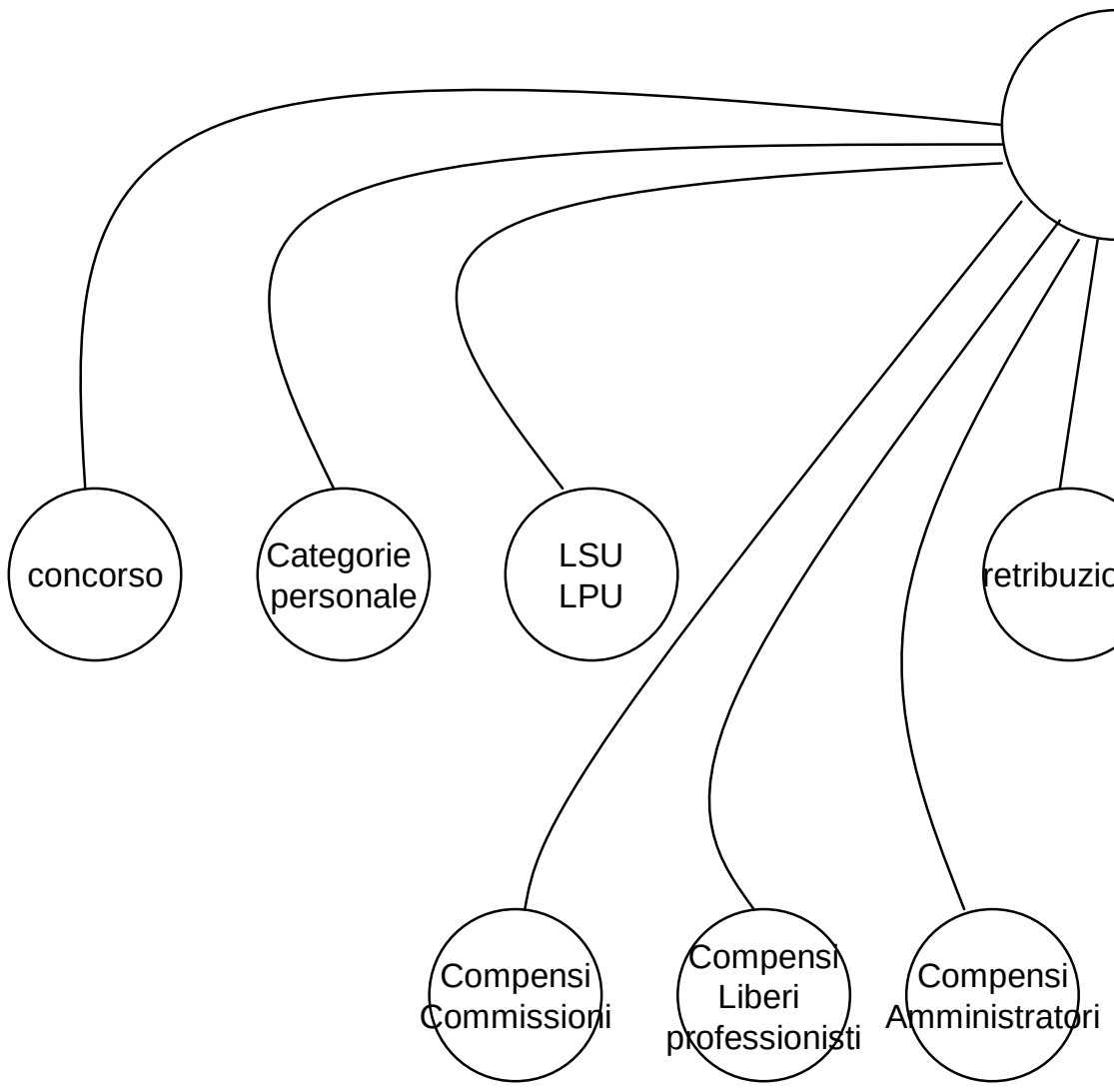

**Passo 3: Generalizzazione**

La generalizzazione consiste nella definizione di macro-categorie generali che raggruppano un insieme di classi di interesse. In particolare si osserva che:

- è possibile definire una classe *misure contro la disoccupazione* quale generalizzazione delle classi *LSU/LPU, esodo* e *trattenimento in servizio*;
- è possibile definire una classe *trattamento economico* quale generalizzazione delle classi *compensi alle commissioni*, *compensi a liberi professionisti* e *compensi agli amministratori*;
- è possibile definire una classe *altre voci di retribuzione* quale generalizzazione delle classi *rimborso spese* e *straordinari*;
- è possibile definire una classe *controversie del lavoro* quale generalizzazione della classe *scioperi*;
- è possibile definire una classe *ordinamento del personale* quale generalizzazione delle classi *concorso* e *categorie del personale*;
- è possibile definire una classe *tipi di lavoro* quale generalizzazione della classe *consulenti*;
- è possibile definire una classe *giurisdizione e normativa del lavoro* quale generalizzazione delle classi *controversie del lavoro, ordinamento del personale* e *tipi di lavoro*;
- la classe *retribuzione* può essere considerata come una generalizzazione della classe *altre voci di retribuzione*.

A valle del processo di generalizzazione la tipologia modificata si presenta come nella figura seguente.

- pubblico impiego
  - giurisdizione e normativa del lavoro
    - controversie di lavoro
      - sciopero
    - ordinamento del personale
      - concorso
      - categorie del personale
    - tipi di lavoro
      - consulenti liberi professionisti e lavoratori autonomi
  - misure contro la disoccupazione
    - lpu ed lsu
    - prepensionamento ed esodo volontario
    - trattenimento in servizio
  - trattamento economico
    - compensi a consulenti liberi professionisti e lavoratori autonomi
    - compensi agli amministratori
    - compensi commissioni gare e concorsi
  - retribuzione
    - altre voci di retribuzione
      - rimborso spese
      - straordinari

## Passo 4: Specializzazione

La fase di specializzazione consiste nell’aggiunta di categorie di interesse come nodi “figli” delle categorie generali, allo scopo di individuare con maggiore precisione il contenuto dei documenti di riferimento.

Nel caso di specie sono previste numerose operazioni di specializzazione, dovute alla grande varietà di contenuti nell’ambito del dominio descritto. Il risultato finale è rappresentato nella figura seguente.

- pubblico impiego
  - addestramento e aggiornamento professionale

- giurisdizione e normativa del lavoro
  - assenze dal lavoro
    - aspettativa dal servizio
    - congedo per cure
    - ferie
    - permessi retribuiti
  - contratti di lavoro
    - contratti collettivi di lavoro
    - contratti di lavoro a tempo determinato
    - contratti di lavoro a tempo indeterminato
    - contratti individuali di lavoro
  - controversie di lavoro
    - processo del lavoro
    - sciopero
  - dimissioni
  - orario di lavoro
  - ordinamento del personale
    - assunzione al lavoro
    - concorso
      - concorsi esterno
      - concorso interno
    - disciplina del rapporto di lavoro
    - inquadramento di personale
    - personale per categorie
      - altro personale
      - categoria a
      - categoria b
      - categoria c
      - categoria d
      - dirigenti
    - sviluppo di carriera
  - parti sociali
    - collegi e ordini

professionali

        - organizzazioni dei datori di lavoro
        - sindacati
    - pensionamento
    - tipi di lavoro
        - consulenti liberi professionisti e lavoratori autonomi
        - lavoratori dipendenti
        - lavoratori stagionali e temporanei
        - lavoro atipico
- misure contro la disoccupazione
    - lpu ed lsu
    - prepensionamento ed esodo volontario
    - trattenimento in servizio
- mobilita dei lavoratori
- trattamento economico
    - compensi a consulenti liberi professionisti e lavoratori autonomi
    - compensi agli amministratori
    - compensi commissioni gare e concorsi
    - retribuzione
        - altre voci di retribuzione
            - assegni familiari
            - buoni pasto
            - premi di produzione
            - progetto obiettivo
            - rimborso spese
            - straordinari
        - competenze
        - trattenute
    - trattamento di fine rapporto
- tutela dei lavoratori
    - assicurazioni sul lavoro
    - assistenza sanitaria
    - infortuni sul lavoro e cause di servizio

- lavoratori madri e padri
- sicurezza nel lavoro

## 5.3 Regole di classificazione

Le regole di classificazione definite sono di due tipi: (i) regola di default *padre-figlio* e (ii) regole basate su nGrammi positivi e negativi. La tabella seguente definisce gli nGrammi positivi e negativi per ogni categoria dell'ontologia.

| CATEGORIA | Positivi | Negativi |
|---|---|---|
| addestramento e aggiornamento professionale | • bilancio n 1040<br>• spese formazione aggiornamento<br>• spese formazione aggiornament<br>• spese formaz<br>• capitolo n 1040<br>• capitolo bilancio 1040<br>• capitolo 1040<br>• cap1040<br>• cap bilancio 1040<br>• cap 1040 | |
| altre voci di retribuzione | • direttivo rag<br>• competenze maturate<br>• aumenti contrattuali<br>• attuazione stesso<br>• approvare schema convenzione, che successive<br>• annui<br>• anno duemilauno giorno ventotto mese<br>• accessorio attesa propria competenza art<br>• 179 16 2, capitolo 2164 comp | |
| aspettativa dal servizio | • chiesto concessione, opportuno procedere<br>• giorni congedo, vita | |
| assegni familiari | • assegni familiari<br>• ciascuno essi, situazione economica | |
| assicurazioni sul lavoro | • giovent sollievo disoccupazione<br>• iniziative favore giovent<br>• malattie professionali | • capitolo 1903 |
| assunzione al lavoro | • assunzione prova | |
| categoria d | • 7 qualifica funzionale<br>• settima qualifica funzionale<br>• categoria d1<br>• categoria d2<br>• categoria d3<br>• categoria d4 | |
| compensi a consulenti liberi professionisti e lavoratori autonomi | • fiducia<br>• garritano<br>• liberi, pubblici<br>• sentenze<br>• unanimita<br>• daniele<br>• affidata<br>• 0982, inoltrato<br>• 1058<br>• 1076<br>• 1206, 153<br>• add | • abbia |
| compensi agli amministratori | • indennit presenza<br>• indennit funzione presidente<br>• indennit carica<br>• gettoni presenza<br>• gettone presenza<br>• consiglieri comunali, gettoni<br>• cap 1013<br>• amministratori, gettoni | |
| compensi commissioni gare e concorsi | • capitolo n 1041, commissione giudicatrice<br>• cap 1041, commissione giudicatrice | |
| concorsi esterno | • progetto, pubblicit<br>• nomina, pubblico<br>• avviso selezione pubblica<br>• assunzione prova medesimo, deliberazione immediatamente<br>• 35 legge 142 90 unanimit, diffusione | • letto<br>• avviso, retribuzioni oneri riflessi |
| concorso interno | • concorso interno<br>• selezione interna orizzontale<br>• selezione interna verticale | • render vacante, seguito concorso<br>• liquidazione compenso |
| consulenti liberi professionisti e lavoratori autonomi | • professionista esterno<br>• procedere nomina professionista<br>• convenzione, professionista<br>• convenzione incarico professionale<br>• convenzione conferimento incarico<br>• consulente esterno<br>• conferimento incarico professionale<br>• collaborazione coordinata continuativa, incarico<br>• alto contenuto professionale | • avviso |
| contratti collettivi di lavoro | • attraverso progressione verticale, banca<br>• annui, che determinano<br>• 4 nuovo, 3 settore affari<br>• 12 ore giornaliere | |
| contratti di lavoro | • 1938 che contratto | |
| infortuni sul lavoro e cause di servizio | • cause servizio<br>• causa servizio | |
| inquadramento di personale | • mutamento | |
| | mansione<br>• mutamento qualifica<br>• reinquadramento personale | |
| lavoratori dipendenti | • dipendenti, lavoro straordinario<br>• mercoled<br>• procedere, trattamento accessorio<br>• proprio mezzo<br>• come liquida favore personale<br>• che dipendente ruolo ente sig<br>• 000 quale, protocollo generale ente data<br>• 1024, servizi generali<br>• 107 109 d lgs, operazione<br>• 193 12 6<br>• 1972 qualifica responsabile, copie responsabile servizio segreteria che<br>• a tempo parziale<br>• approvato piano lavoro proposto ufficio | • 000 000<br>• Convenzione<br>• permesso<br>• entro 30 |
| lavoratori madri e padri | • come sopradescritta a, risulta documentazione presentata<br>• anno vita bambino<br>• 1999 n, 2001 oggetto concessione<br>• 06 2003, 2003 30 | |
| lavoratori stagionali e temporanei | • tempo determinato<br>• assunzioni temporanee stagionali<br>• assunzione temporanea<br>• assunzione diretta, unit lavorative<br>• assunzione diretta, unit lavorativa<br>• assunzione diretta, | |
| | durata giorni | |
| lavoro atipico | • 1280, fornitura lavoro temporaneo<br>• compilati | |
| lpu ed lsu | • convenzione n<br>• lsu<br>• socialmente utili | • presso inail<br>• 5 95 |
| misure contro la disoccupazione | • disoccupazione, voti<br>• atto forma | • n richiamata |
| orario di lavoro | • seguente orario<br>• n 2 ore<br>• lavoro tempo<br>• 53 2000 art, figlio<br>• 27 4 1972, cgil<br>• 00 ore, 10 legge n | |
| ordinamento del personale | • organismo<br>• 1972, distacco<br>• 108, aggravio | |
| Pensionamento | • collocamento pensione, limiti eta, anno et limite | |
| permessi retribuiti | • 10 legge n<br>• permessi riducono ferie | |
| prepensionamento ed esodo volontario | • esodo volontario dipendente | |
| progetto obiettivo | • 357, attestare<br>• 352, rendimento<br>• 2006, avente | |
| Retribuzione | • 000x<br>• decurtazioni<br>• ultime | |
| rimborso spese | • onere a carico<br>• persona dr<br>• missione<br>• economo comunale, esecutivo gestione<br>• assessori personale<br>• a titolo indennit<br>• 1024<br>• 00 a saldo | • svolgono<br>• nominativi<br>• 4 5, indennit missione |
| Sciopero | • aderito sciopero | |
| Sindacati | • aderito sciopero | |
| Straordinari | • 31 oneri riflessi<br>• 1873, integrare<br>• occasione festeggiamenti | • presenza<br>• paola 20<br>• effettuate personale |
| | • lavoro straordinario<br>• 2165 corrente bilancio<br>• contratto, periodo sopra | • 833<br>• 1 gennaio |
| sviluppo di carriera | • categoria d1 | |
| trattamento economico | • operata inps conto<br>• 1873 2 corrente, 2000 rilevato<br>• 18 96, allegato suddetta<br>• 12 ore giornaliere | |
| trattenimento in servizio | • 468 97<br>• 503 92 preso, prosecuzione | |
| Trattenute | • 1938 che contratto<br>• aderito sciopero | |

Per ragioni di sintesi si mostrano di seguito le regole limitatamente alla sola categoria “concorso interno”.

```
positive("concorso interno",IdDoc) :- twogram(IdDoc,"concorso
interno",_,_,_).

negative("concorso interno",IdDoc) :- twogram(IdDoc,"render
vacante",_,_,_),  twogram(IdDoc,"seguito concorso",_,_,_).

negative("concorso interno",IdDoc) :-twogram(IdDoc,"liquidazione
compenso",_,_,_).

positive("concorso interno",IdDoc) :-threegram(IdDoc,"selezione interna
verticale",_,_,_).

positive("concorso interno",IdDoc) :-threegram(IdDoc,"selezione interna
orizzontale",_,_,_).

success("concorso interno",IdDoc,100,100,100) :- positive("concorso
interno",IdDoc), not negative("concorso interno",IdDoc).
```

**BIBLIOGRAFIA**